\documentclass[final]{aipproc}
\layoutstyle{8x11double}
\usepackage{epstopdf}

\usepackage{graphicx}

\newcommand{\onepion}{one pion\ }
\newcommand{\onenucleon}{one nucleon\ }

\newcommand{\GeV}{\; \mathrm{GeV}}

\newcommand{\dd}{\mathrm{d}}

\newcommand{\Minerva}{Miner$\nu$a\ }
\newcommand{\Nova}{No$\nu$A\ }

\begin{document}

\title{GiBUU and Shallow Inelastic Scattering}
\author{O. Lalakulich}{
  address={Institut f\"ur Theoretische Physik, Universit\"at Giessen, Germany}
}
\author{U. Mosel}{}

\begin{abstract}
In this talk we shortly describe the physics contents of the GiBUU transport code,
used to describe lepton scattering off nuclei. Particular attention will be given to validation
of the GiBUU in pion-, electron- and photon-induced reactions, which serve as a benchmark for
neutrino-induced ones. We mainly concentrate on those properties of benchmark reactions, which are
relevant to the region of Shallow Inelastic Scattering (SIS). Our results in this region are presented
for integrated and differential cross sections. Comparison with recent MINOS inclusive data,
as well as predictions for the differential cross sections measurable in \Minerva and \Nova
experiments are made.
\end{abstract}

\date{\today}

\maketitle

\section{Introduction}

An inevitable feature of all neutrino experiments is that they have to deal with wide-energy band beams.
For the \Nova and MINOS experiments
(as well as \Minerva and ArgoNeut, which are using the same flux as MINOS)
this energy region starts around 1 GeV, where QE scattering and Delta production dominate
the neutrino cross section and ranges up to several tens of GeV, where DIS dominates.
The peak of the energy distribution at only a few GeV causes higher-lying resonances and the
transition region between resonances and DIS still to be important.

The superposition of quite different reaction mechanisms in these experiments complicates the
reconstruction of the incoming neutrino energy since, e.g., quasielastic (QE) scattering can not cleanly
be identified. This then affects among others the extraction of neutrino oscillation parameters \cite{Lalakulich:2012hs}.
It is therefore essential to describe \emph{all} the relevant reaction mechanisms in a reliable way.
With an emphasis on this aspect we discuss first the GiBUU implementation of transport theory and its various
verifications. In a second part of this paper we then go on to a discussion of SIS processes on nuclear targets.


\section{GiBUU}

The GiBUU model has been developed
as a transport model for nucleon-, nucleus-,  pion-, and electron-induced collisions from
some MeV up to tens of GeV. Several years ago neutrino-induced interactions were
also implemented for the energies up to a about 2 GeV \cite{Leitner:2006ww,Leitner:2006sp}
and, recently, the GiBUU code was extended to describe also the DIS processes for neutrino reactions.

Thus, with GiBUU it is possible to study various, quite different reactions
on  nuclei within a unified framework~\cite{Buss:2011mx}.
Relevant for the present investigation or neutrino interactions with nuclei is also the fact that the method and code have been
widely tested for photon-induced as well as for electron-induced reactions in the
energy regime from a few hundred MeV to 200 GeV
\cite{Effenberger:1999jc,Krusche:2004uw,Buss:2007ar,Gallmeister:2007an,Leitner:2008ue}.
In this section we present the main steps in the GiBUU code and how the input physics
is tested.

\subsection{Initialize nucleus}

\begin{figure}[!bht]
\includegraphics[width=\columnwidth, height=6cm]{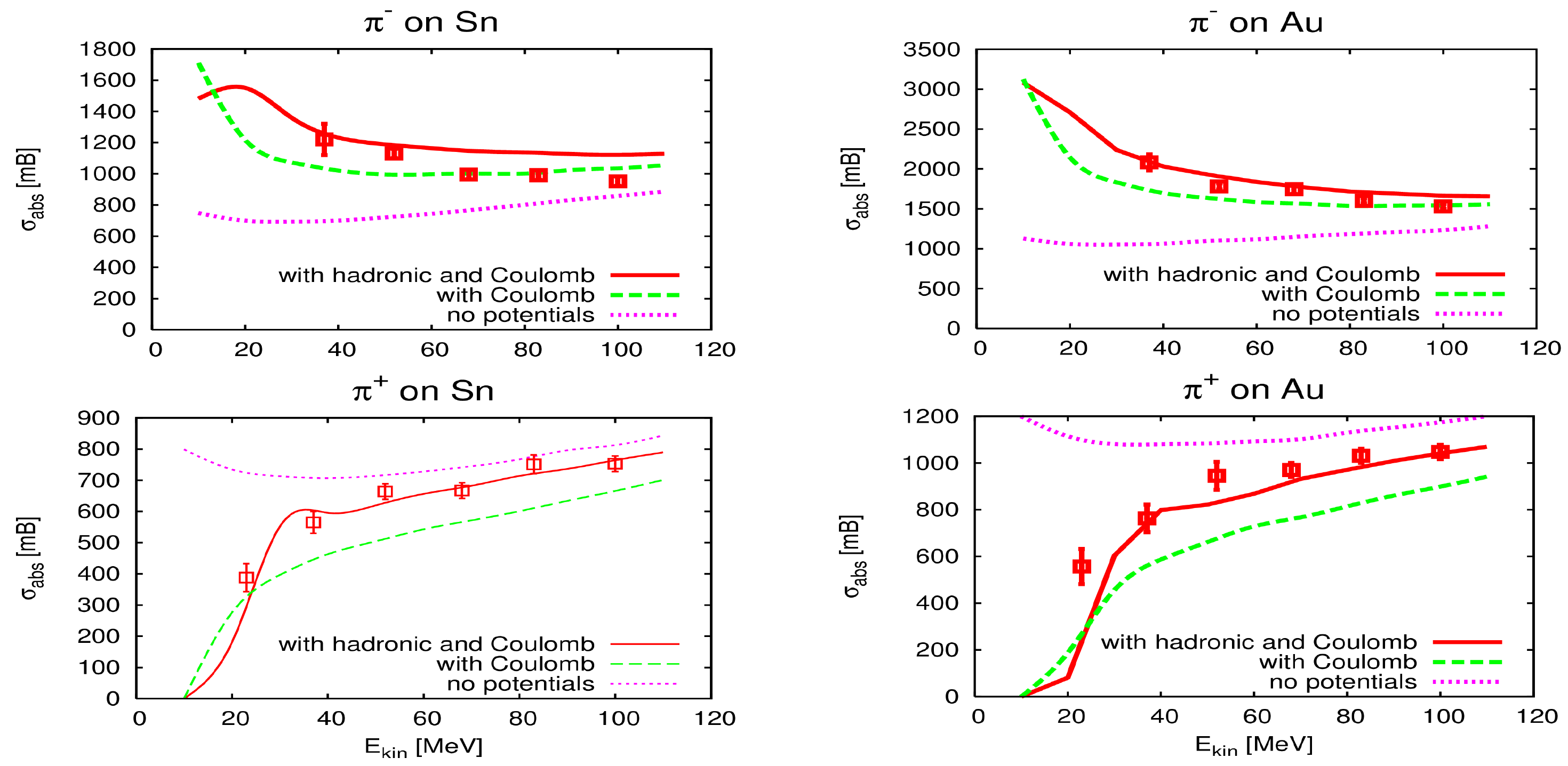}
\caption{Cross section  for $\pi^+$ and $\pi^-$ absorption on Sn and Au versus kinetic energy of the
incomig pions. The necessity to include nuclear and Coulomb potentials is clearly demonstrated.
Calculations without  including the potential (magenta dotted lines) underestimate experimental
data for $\pi^-$ and overestimate  the data for $\pi^+$ reactions. Including nuclear potential
(dashed green lines) significantly  improve the agreement.
Including further the Coulomb potential (red solid lines) brings the curves to the data.}
\label{fig:pionAbsorption}
\end{figure}

At the fist step of GiBUU simulation for any reaction the struck nucleus is to be initialized.
GiBUU describes it as a collection of off-shell nucleons.
Each nucleon is bound in a mean-field potential, which on average describes the many-body interactions with the other
nucleons. This potential is parameterized as a sum of a Skyrme term depending only on density, and
a momentum--dependent contribution.
The phase space density of nucleons is treated within a local Thomas-Fermi approximation.
At each space point the nucleon momentum distribution is given by a Fermi sphere, whose radius
in  momentum space is determined by the local Fermi momentum which depends on the nucleon
density.  Within this picture, contrary to the  Fermi gas model with constant Fermi momentum
(the global Fermi gas model), the nucleon position and momentum are correlated. This leads to
a smoother momentum distribution with somewhat more strength at lower momenta and to smoother
nucleon spectral functions.

The validation of such a description can be checked, for example, with pion absorption reactions.
Fig.~\ref{fig:pionAbsorption} shows the absorption cross section for negatively and positively charged pion on tin and
gold.

\subsection{Calculate in-medium cross section}

\begin{figure}[htb]
\includegraphics[width=\columnwidth]{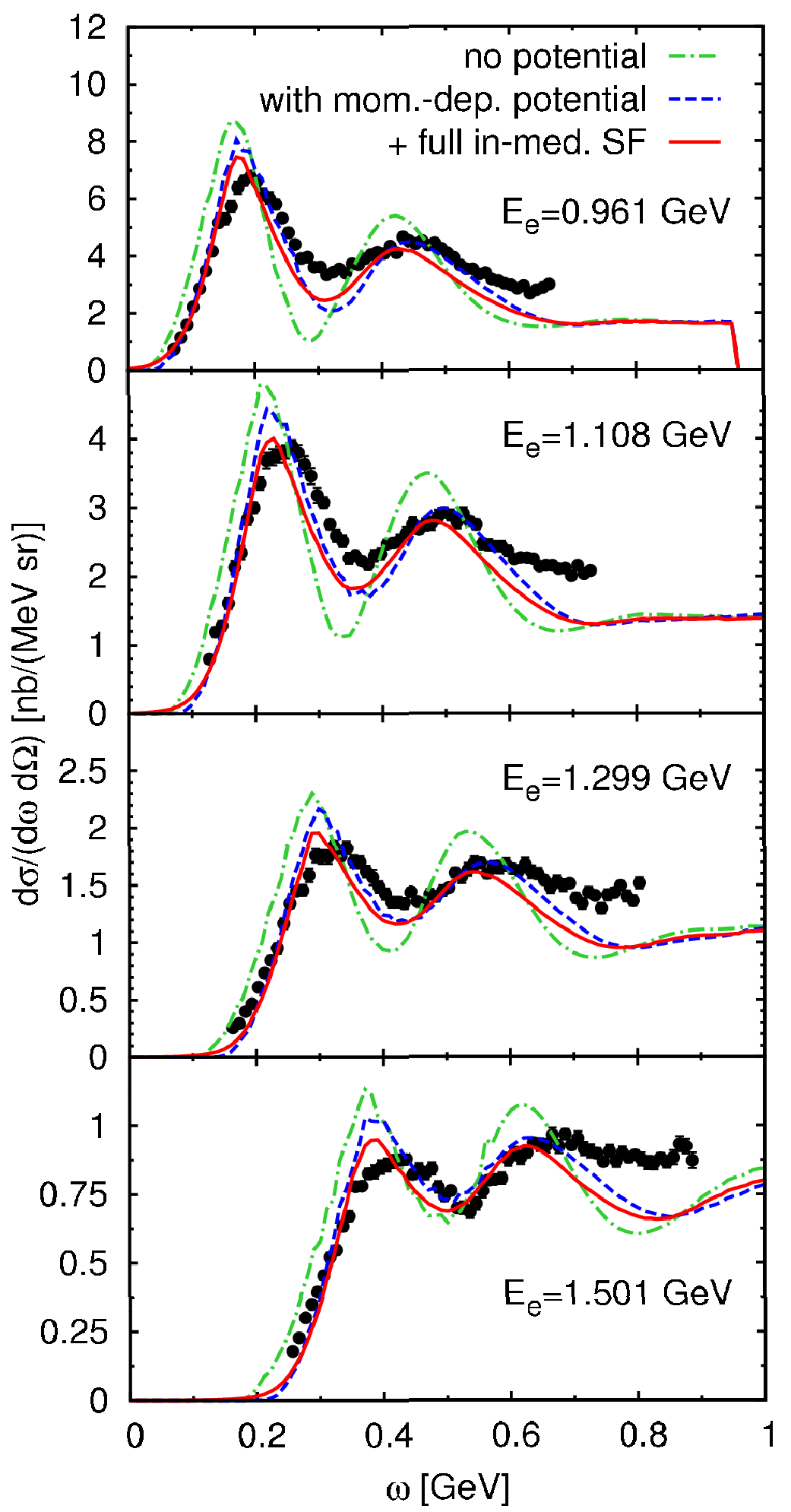}
\caption{Cross section $\dd\sigma/\dd\cos\theta_\mu \dd\omega$ for inclusive electron scattering on carbon.
Curves obtained without including any potential (green dashed curves) overestimate both QE and $\Delta$ peaks
and give a too deep dip in between. Including momentum-dependent nuclear potential in the model (blue short-dashed curves)
significantly improve agreement with the data. Taking into account the full in-medium spectral function
(red solid curves), that is medium-modified width and mass of the baryons brings further improvement.}
\label{fig:inclusiveElectron}
\end{figure}

In view of our purpose to use the model for description of neutrino scattering,
such verification of the model with pion scattering is necessary but not sufficient.
Being strongly interacting particles, pions have a small penetration length and thus
mainly interact close to the surface of the nucleus. Electrons and photons,
as well as neutrinos, on the other hand, probe the whole volume of the nucleus.
Validation of the GiBUU model with these probes can be demonstrated on the example of the
inclusive electron scattering off carbon nucleus.
The double differential cross sections $d\sigma/d\cos\theta_\mu d\omega$ are shown in
Fig.~\ref{fig:inclusiveElectron} for various energies of the incoming electron and
fixed scattering angle of $\theta_\mu=37.5^\circ$ versus the energy transfer $\omega$.
The GiBUU calculations are compared to the JLab data.
This figure again demonstrate the sensitivity of the results to the choice of the nuclear
potential.

\begin{figure*}[htb]
\includegraphics[width=0.7\textwidth]{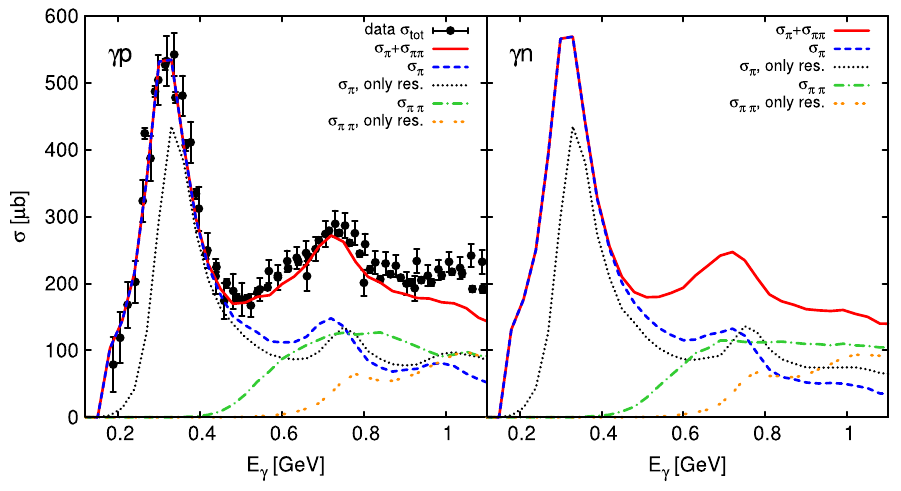}
\caption{Photoabsorption cross section on proton and neutron targets versus photon energy.
To reach agreement with the data in the $\Delta$ peak region, both resonant (dotted black curve)
and non-resonant pion production are needed. The latter is tuned in such a way, that
their sum (short-dashed blue curve) fits the data. At higher energies both resonant (double-dotted
yellow curve)  and non-resonant 2-pion production are needed. The latter is tuned in such a way, that their
sum (long-dashed green curve) fits the data. These calculations do not contain any $2\pi$ or DIS contributions which contribute at the higher
energy transfers.}
\label{fig:photoabsorption}
\end{figure*}

Another important feature of the GiBUU model is that it incorporates not only resonance
but also background contributions. Its necessity can be clearly validated using the photoabsorption data.
Fig.~\ref{fig:photoabsorption} shows the calculated photoabsorption cross section for
proton and neutron targets.  Agreement with the data
can only be reached when all reaction channels are taken into account.
Those are  both resonance and background 1-pion production in the $\Delta-$region, up to photon energies
of about 0.4 GeV.  The threshold for the production of high-mass resonances, able to decay to 2 pions,
open only at $E_\gamma > 0.6 \GeV$. In this energy region the second resonance peak at $E_\gamma
\approx 0.7 \GeV$ is clearly visible. In the region of 0.4-0.6 GeV, that is below the resonance production
threshold, the non-resonant 2-pion production is indispensable.
One should keep this in mind, when considering neutrinoproduction: both resonant and
non-resonant (background, direct pion production)  processes must be taken into account.

\begin{figure*}[htb]
\includegraphics[width=0.7\textwidth]{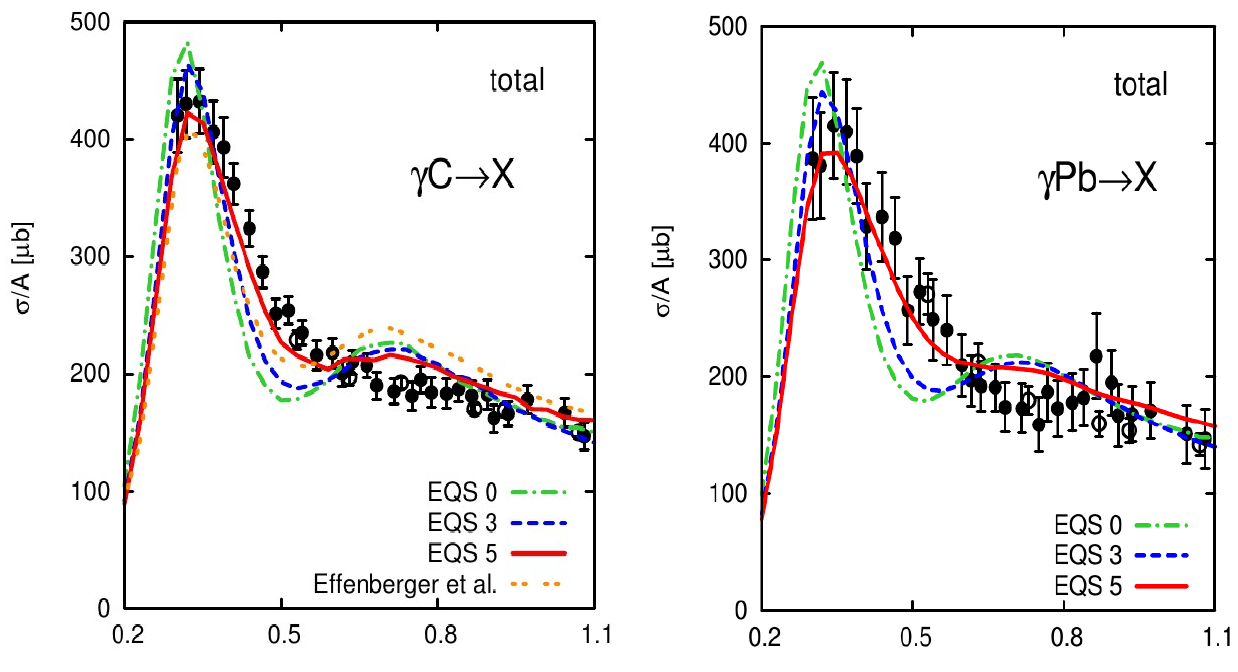}
\caption{Photoabsorption cross section on carbon and lead targets versus photon energy.
Calculation without nuclear potential (dashed green curves, labeled ``EQS 0'')
clearly shows the second peak not present in the data. Including nuclear potential
without momentum dependence, the so-called soft potential (short-dashed blue curve, labeled ``EQS 3'')
slightly increases the cross section in the dip region. Using the momentum-dependent potential,
medium momentum-dependent in particular (red solid curves, labeled ``EQS 5'') noticeably increases the
cross section in the dip region and slightly decreases it in the second peak region, thus
improving agreement with the data.}
\label{fig:photoabsorption-nucleus}
\end{figure*}

When going from protons to nuclear targets, we face a surprising problem that is not yet
fully clarified in electroproduction theory --- the second resonance peak disappears
in photoabsorption on nuclei. This is demonstrated on the example of carbon and lead
targets, as shown in Fig.~\ref{fig:photoabsorption-nucleus}. Partly this effect is due
to the medium modification of the resonance properties, partly due to the presence of the nuclear
potential. Including the momentum-dependent nuclear potential in the model significantly improves
the agreement with the data, as compared to the calculations without nuclear potential. The deficiency of
the calculated cross section at the high-energy side of the $\Delta$ peak could be an indication for missing
$2p2h1\pi$ processes.

\subsection{Propagate outgoing hadrons throughout the nucleus}

A remarkable feature of the GiBUU, which distinguishes it from other neutrino event generators, is the implementation of sophisticated treatment of final state interactions.
Hadrons produced in the initial interaction act inside the nucleus,
can rescatter off another bound nucleon, changing their energy, and/or producing additional mesons
and/or knocking-out this nucleon.
Pions, that were originally produced through a weak excitation of $\Delta$ resonance, can be absorbed in the nucleus
or converted to other mesons. Thus, the inevitable presence of the FSI washes out the true origin of the event and makes an
experimentally observed signal different from what one would expect for the  scattering on a free nucleon. FSI can decrease the cross sections as well as significantly modify
the shapes of the final particle spectra.

The restoration of the true cross section of the process of interest, therefore,
cannot be achieved by pure experimental means and crucially relies on theoretical modeling
implemented in the event generators. For a reliable interpretation of experiments one thus
needs a model which provides a realistic description of both initial neutrino-nucleus interaction
and the final state interaction of the produced hadrons.

\begin{figure}[htb]
\includegraphics[width=\columnwidth]{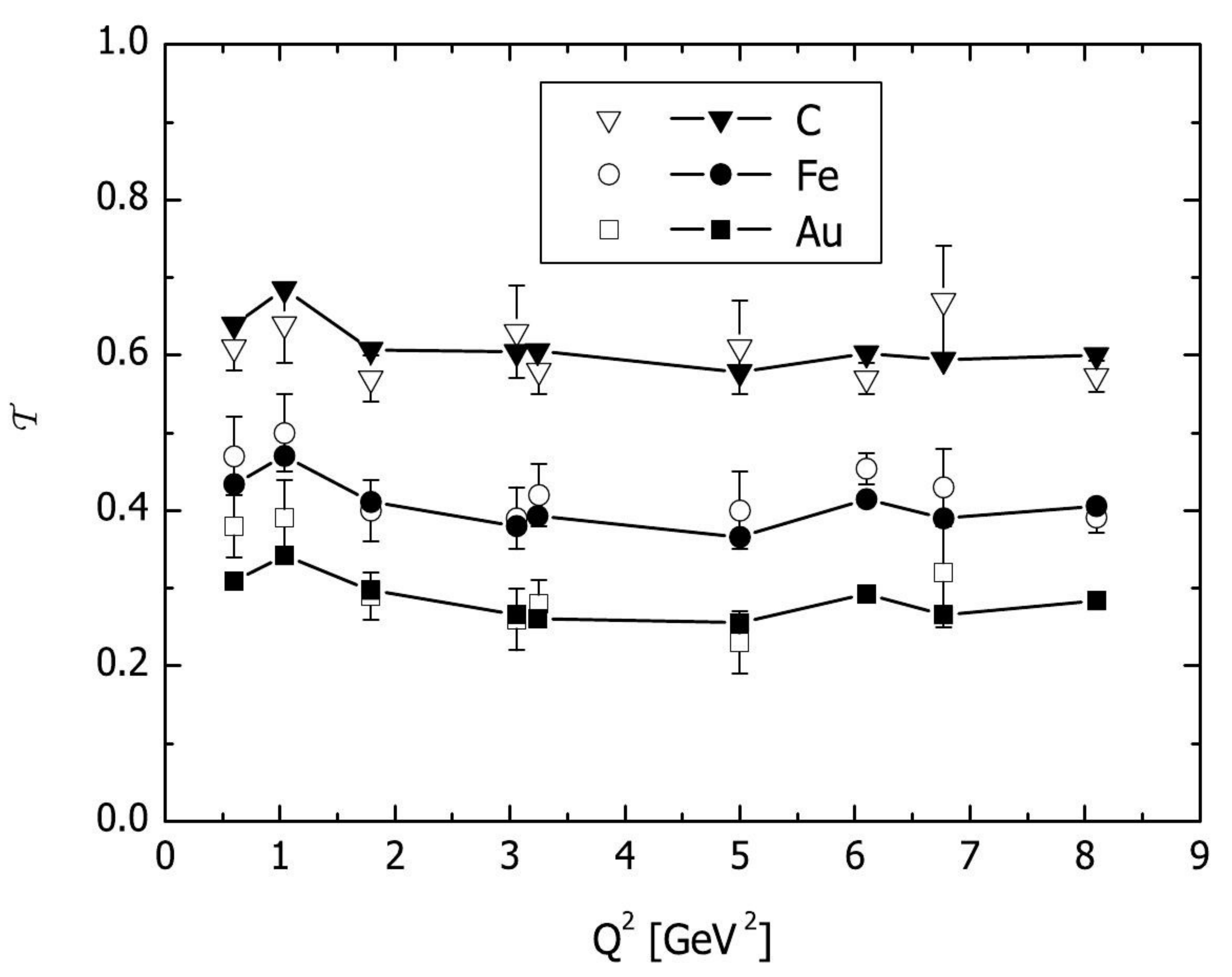}
\caption{Transparency in electron scattering $A(e,e' p)$ off carbon, iron and gold versus the
squared momentum transferred. Experimental data (filled symbols) are perfectly described by the
GiBUU model (open symbols  connected by lines).}
\label{fig:transparencyElectron}
\end{figure}

The existing neutrino event generators describe FSI with the cascade model, mainly because it is
relatively easy to implement and the simulation is not time-consuming.
In GiBUU, on the contrary, FSI are implemented by solving the semi-classical Boltzmann-Uehling-Uhlenbeck
(BUU) equation. It describes the dynamical evolution of the phase space density for
each particle species under the influence of the mean field potential, introduced in the description of
the initial nucleus state. Equations for various particle species are coupled through this mean field and
also through the collision term. This term explicitly accounts for changes in
the phase space density caused by elastic and inelastic collisions between particles.
For a more detailed discussion of FSI see Ref.~\cite{Buss:2011mx}.
Thus, with 61 baryons and 21 mesons included in the model, we are solving the system of 83 coupled
differential equations. The price to pay for this realistic description is computer time for the
calculations.

\begin{figure}[htb]
\includegraphics[width=\columnwidth]{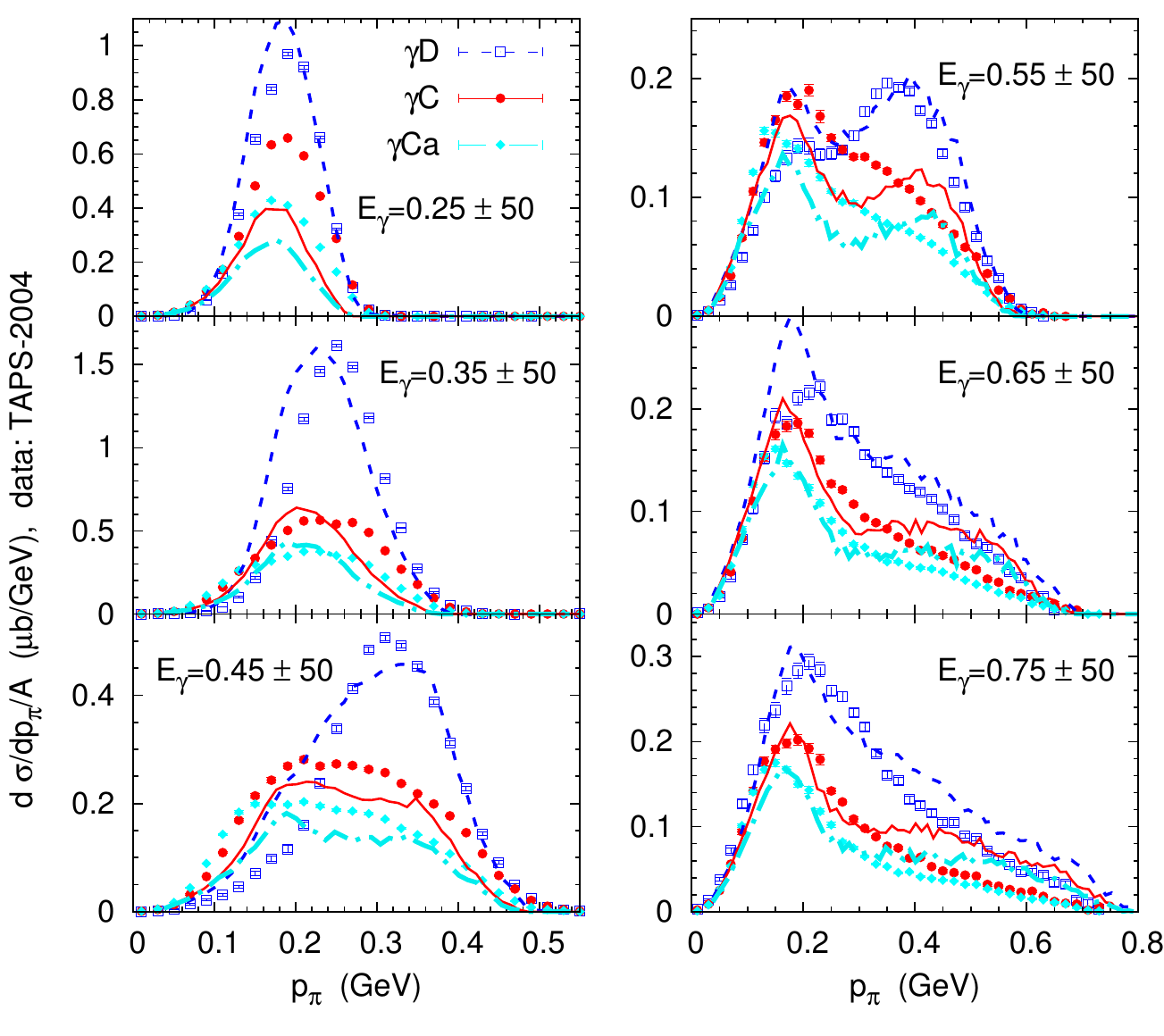}
\caption{Momentum distributions of the outgoing pions for inclusive $\pi^0$ production in scattering
 of photons of energies $0.2-0.8\GeV$ off D (blue short-dashed curves),  C (red solid curves)
 and Ca (dash-dotted cyan curves) nuclei.
 The data are from \cite{Krusche:2004uw}. They contain coherent pion production,
 which is not included in the calculations; it plays a role only on the low-momentum side of the peak. }
\label{fig:pi0photoproduction}
\end{figure}

Validation of our treatment of FSI can be demonstrated by comparison of the
GiBUU calculations of proton transparency ${\cal{T}}$ in electron scattering off carbon, iron and gold
with the JLab/SLAC data. GiBUU calculations show a very good agreement with the data (see Fig.\ \ref{fig:transparencyElectron}).

Closely related to neutrino-induced reactions is the photoproduction of pions.
On the nucleon (i.e.,\ without final state interactions) these reactions  directly
test the vector part of the pion-production vertex.  On nuclei this reaction tests
the nuclear dynamics of pion propagation throughout the nucleus.
GiBUU describes the dataset for photoproduction of neutral pions~\cite{Krusche:2004uw}
on nuclei for photon energies up to $0.8\GeV$ quite well \cite{LehrDiss:2003}.

As an illustration, that will become relevant for the later discussions,
Fig.~\ref{fig:pi0photoproduction} shows measured and calculated pion momentum distributions
for $\pi^0$ photoproduction off D and C nuclei. The shapes of the experimental distributions
change significantly when going from deuterium (which is nearly equivalent to production before FSI)
to  C (corresponding to production after FSI). The main effect is a strong absorption of pions
around momenta of 0.3 GeV due to the excitation of the $\Delta$ resonance and its subsequent
pionless decay. GiBUU calculations reproduce this behavior and show
a generally good agreement with the  data;
in the calculations the absorption around 0.3 GeV is indeed due to
$N\Delta \to NN$ or $NN\Delta \to NNN$ collisions. The remaining discrepancies between theory
and experiment for C and Ca give an indication of
the systematic errors in the GiBUU calculations.
We also note here that the calculations shown in Fig.~\ref{fig:pi0photoproduction} all use the
collision-broadened width as given by Oset and Salcedo \cite{Oset:1987re}.
Using the free width instead would yield significantly too large cross sections at the peak position.

For the energy region around a few GeV also concepts such as 'formation zone' or 'formation time' become important; they
attempt to model the finite time needed for a struck parton to hadronize into a groundstate hadron. At the lower invariant
masses, in the resonance region, the formation time is exclusively determined by
the inverse resonance width $\tau = 1/\Gamma$. The decay products of the resonance, e.g.\ the pions from $\Delta$ decay, appear
according to an exponential decay law and then start to interact immediately; on the other hand, also before pion emission the $\Delta$ resonance can interact with other nucleons in the nucleus and, for example, can be absorbed through $\Delta N \to N N$, an important pion absorption channel. At higher energies where individual resonances melt into a QCD continuum this life-time of resonances is modeled by a formation time during which the interaction of the produced hadron with the nucleons in the nucleus is reduced. In \cite{Falter:2003di,Falter:2004uc} we have used a reduction of this interaction by a constant factor, depending on the number of leading quarks in the formed hadron, which accounts well for the hadronization experiments in the HERMES energy regime. In \cite{Gallmeister:2007an} we have shown a more sophisticated method that contains a linear rise of the interaction cross section of the new formed hadron with the nucleons in the target nucleus can account for hadronization data over a wide
range of leptonenergies, from about 10 to 200 GeV. We use this scheme, therefore, also for the neutrino reactions in the SIS and DIS regions \cite{Lalakulich:2012gm}.

\subsection{Multiplicities}

\begin{figure}[htb]
\begin{tabular}{c}
\includegraphics[width=0.9\columnwidth]{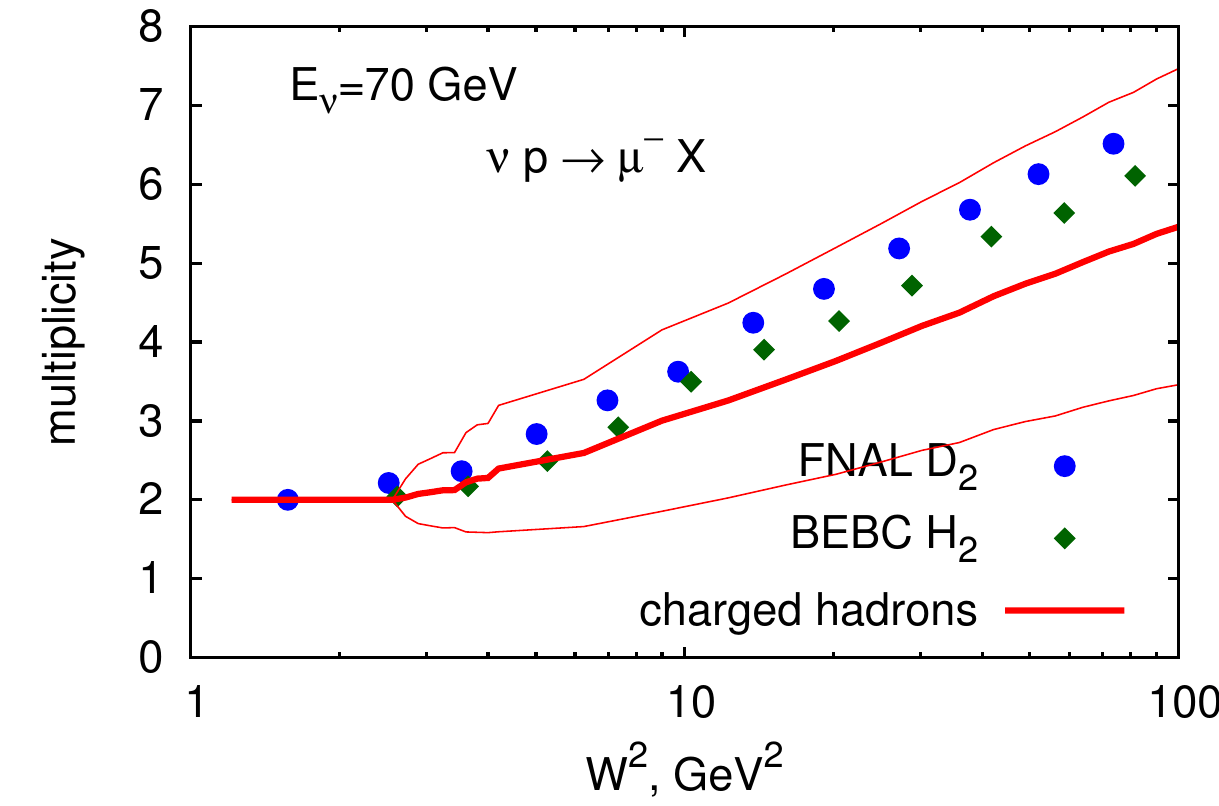}
\\
\includegraphics[width=0.9\columnwidth]{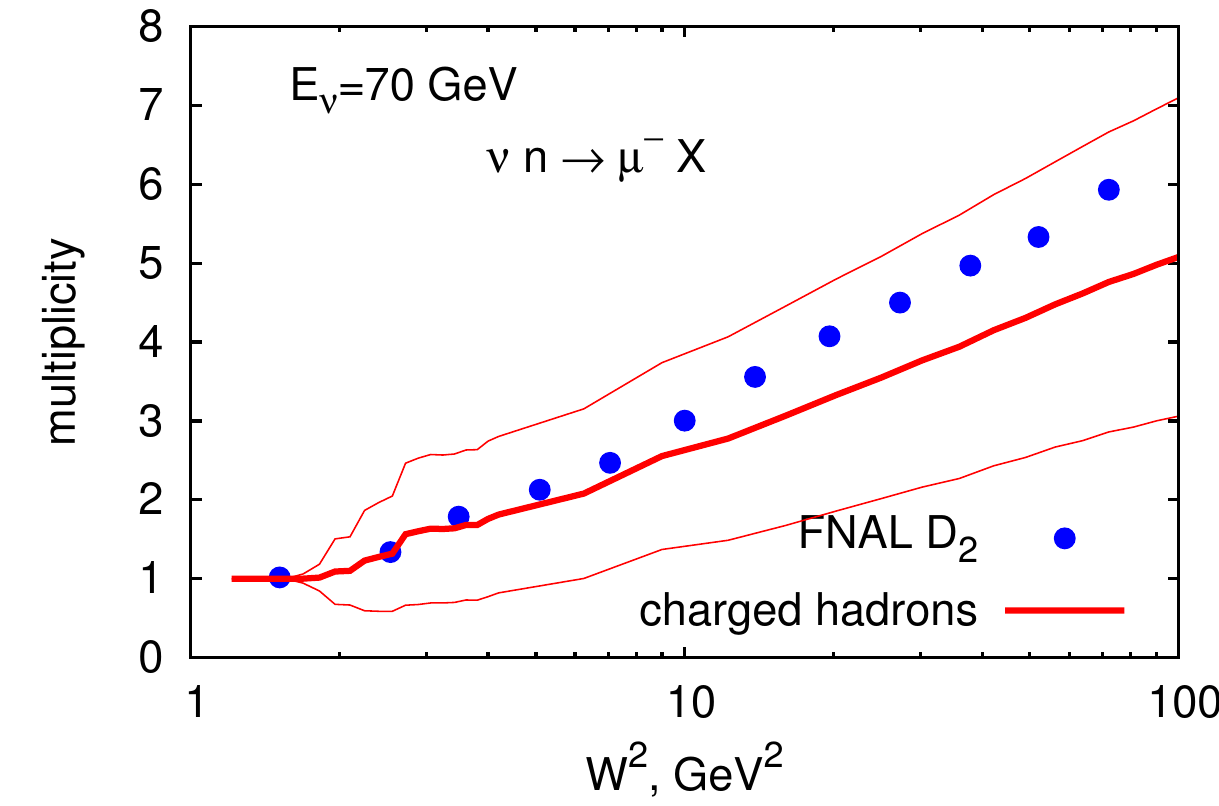}
\end{tabular}
\caption{ Average multiplicities for charged hadron production (red thick solid curves) with
1$\sigma$ confidence intervals (red thin solid curves) for proton (top panel) and neutron
(bottom panel) targets versus the  squared invariant hadron mass $W^2$.
GiBUU calculations are compared with the FNAL (blue circles) data on deuterium target
and BEBC (green diamonds) data on hydrogen target.}
\label{fig:multiplicities-charged-p-n}
\end{figure}
\begin{figure*}[htb]
\begin{tabular}{c}
\includegraphics[width=0.9\columnwidth]{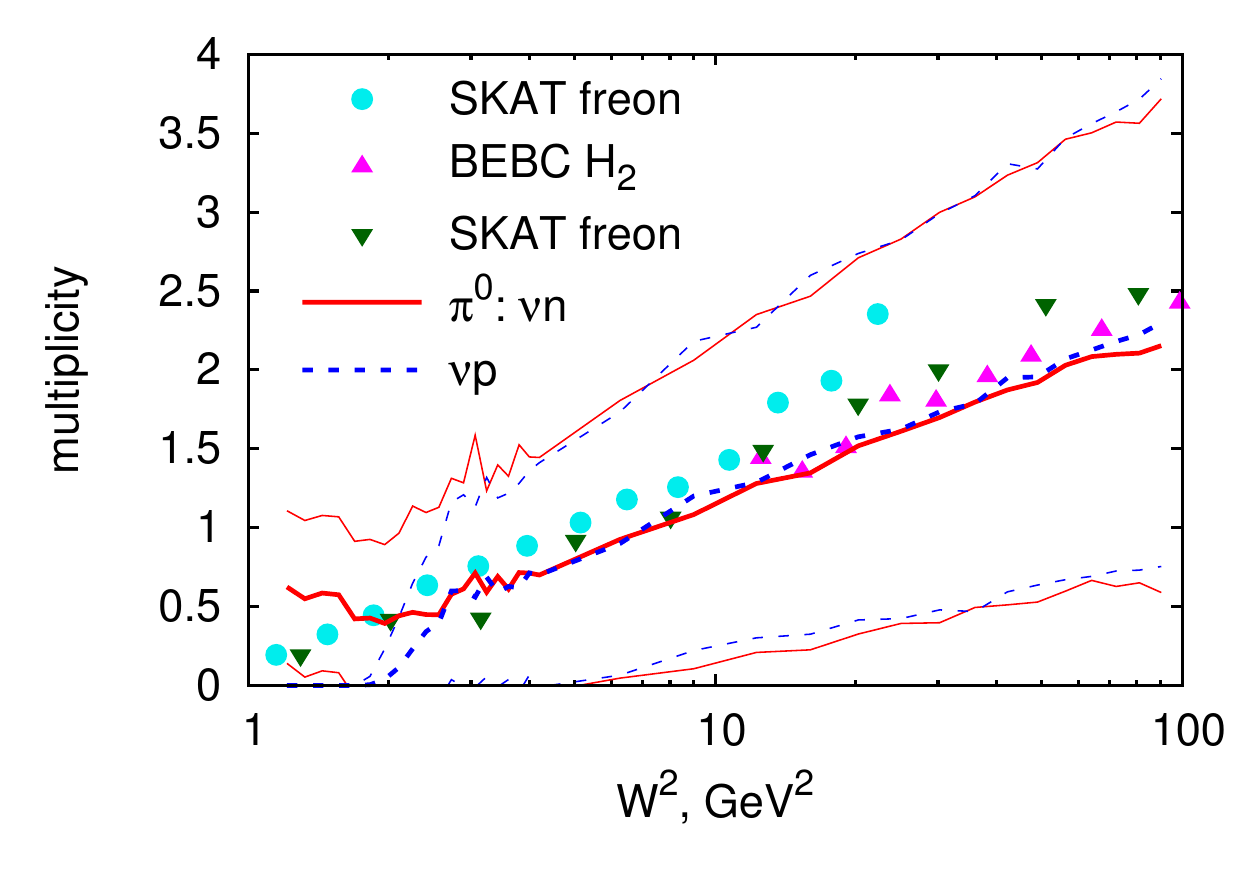}
\\
\includegraphics[width=0.9\columnwidth]{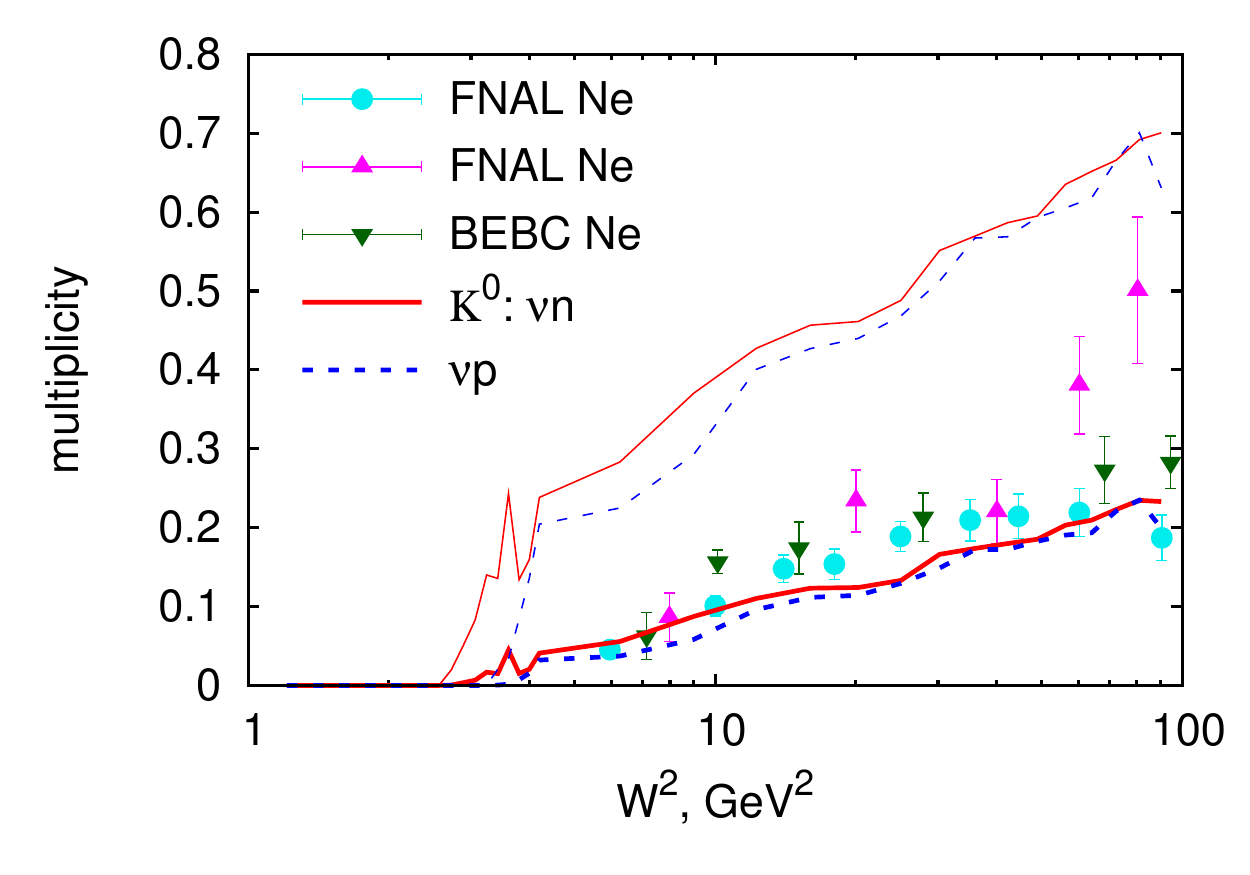}
\end{tabular}
\caption{ Average multiplicities for $\pi^0$ (top panel) and $K^0$ (bottom panel) production versus the
squared invariant hadron mass $W^2$.
GiBUU calculations for proton (blue thick short-dashed curves) and neutron (red thick solid curves) targets
and the corresponding 1$\sigma$ confidence intervals (thin curves) are compared to the SKAT
data for freon and BEBC data for hydrogen.}
\label{fig:multiplicities-pi0-K0}
\end{figure*}

Another important point that should be checked by any neutrino event generator is the
multiplicity of various particles. Such comparison was made first in \cite{Yang:2009zx}
within the GENIE event generator. Here we present the corresponding results from the GiBUU model.

Fig.~\ref{fig:multiplicities-charged-p-n} compares our calculations for
the multiplicities for all charged hadrons with the data available from the old
neutrino experiments.
Fig.~\ref{fig:multiplicities-pi0-K0} shows the multiplicities for production of
neutral pions and kaons. It is interesting to note here that experimentally the
data on proton and nuclear targets nearly coincide. Furthermore, they lie well within a 1-$\sigma$
confidence band (the latter is calculated within GiBUU).

\section{Shallow Inelastic scattering}

Now let us proceed with the neutrino and antineutrino reactions:
\[
\begin{array}{l}
\nu(k^\mu)  N (p^\mu) \to \mu^- (k'{}^\mu) X \ ,
\\[4mm]
\bar\nu(k^\mu)  N (p^\mu) \to \mu^+ (k'{}^\mu) X \ ,
\end{array}
\]
with a bound nucleon inside the nucleus.

\begin{figure}
\begin{tabular}{c}
\includegraphics[width=0.8\columnwidth]{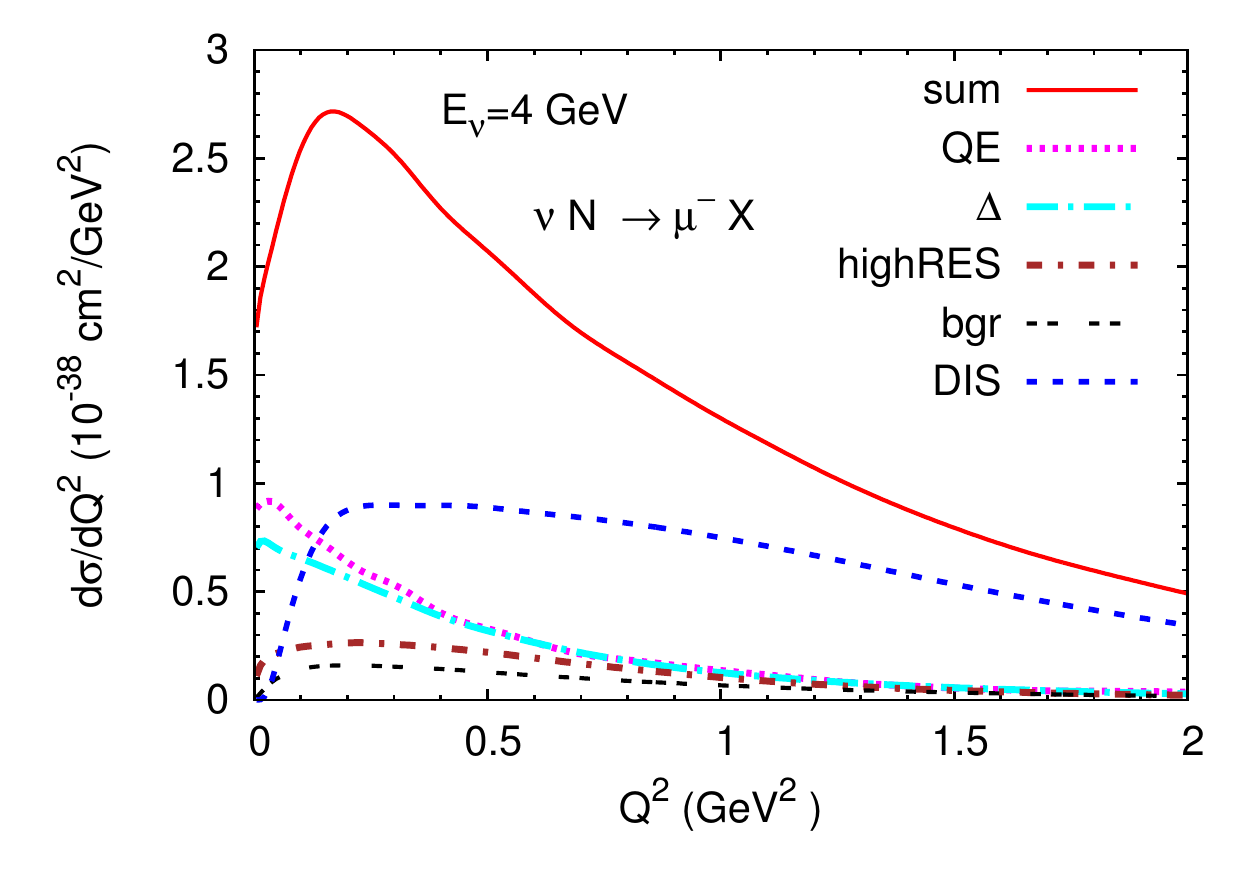}
\\
\includegraphics[width=0.8\columnwidth]{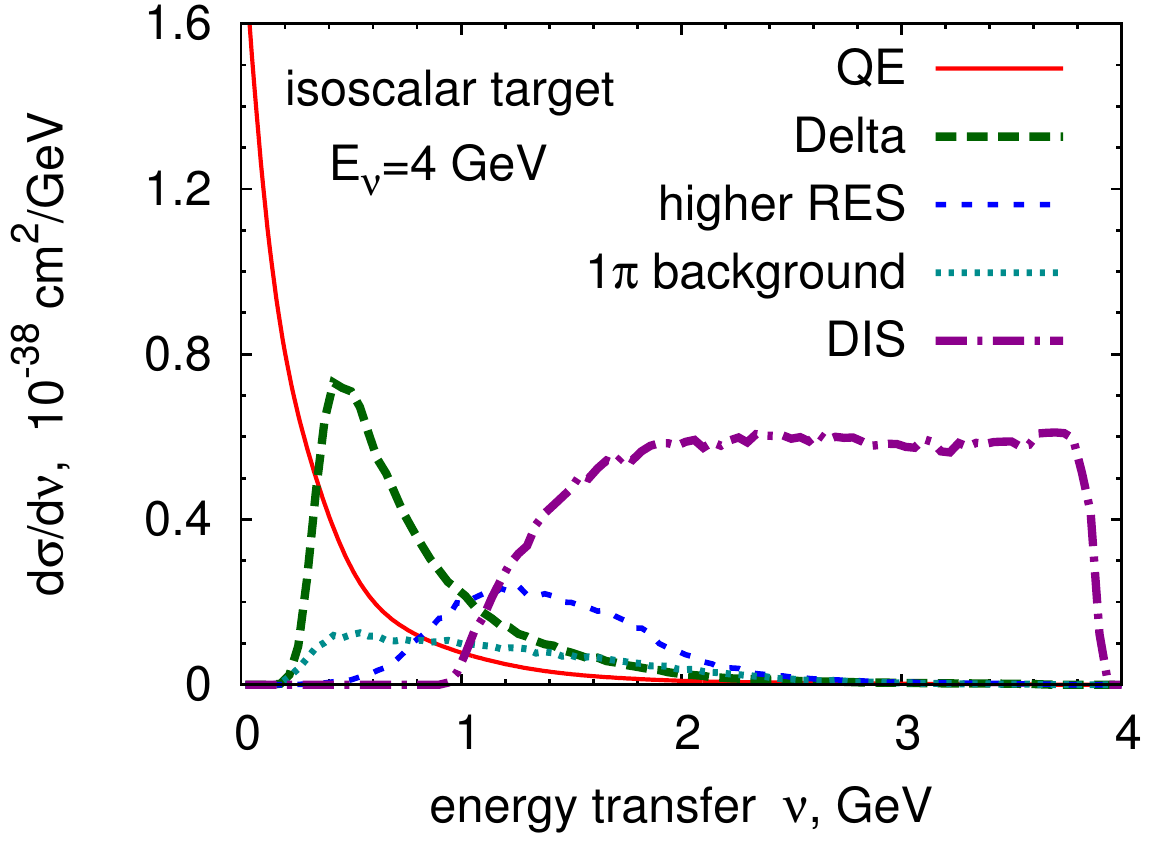}
\end{tabular}
\caption{Cross sections $\dd \sigma/\dd Q^2$ (top panel) and $\dd \sigma/\dd \nu$ (bottom panel)
per nucleon for CC neutrino scattering off an isoscalar target  for $E_{\nu}=4\GeV$.}
\label{fig:isoscalar-dsidQ2-dsidnu}
\end{figure}

\begin{figure}
\begin{tabular}{c}
\includegraphics[width=0.8\columnwidth]{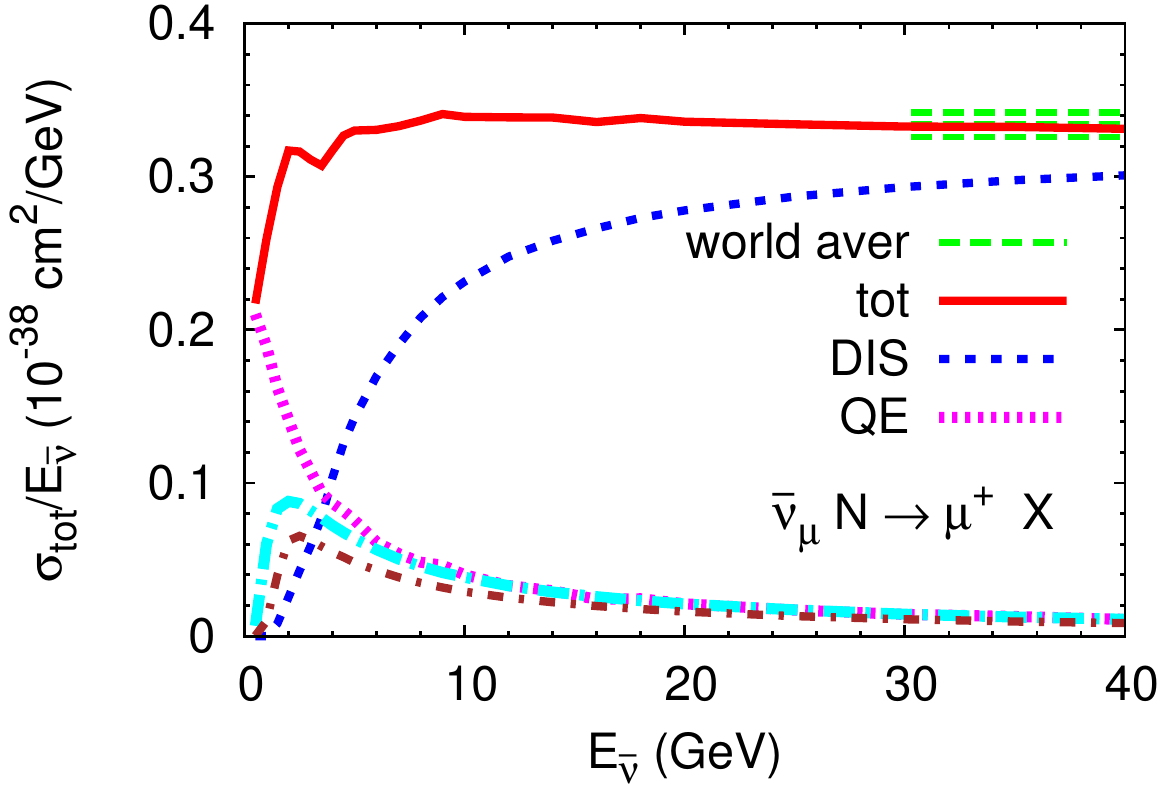}
\\
\includegraphics[width=0.8\columnwidth]{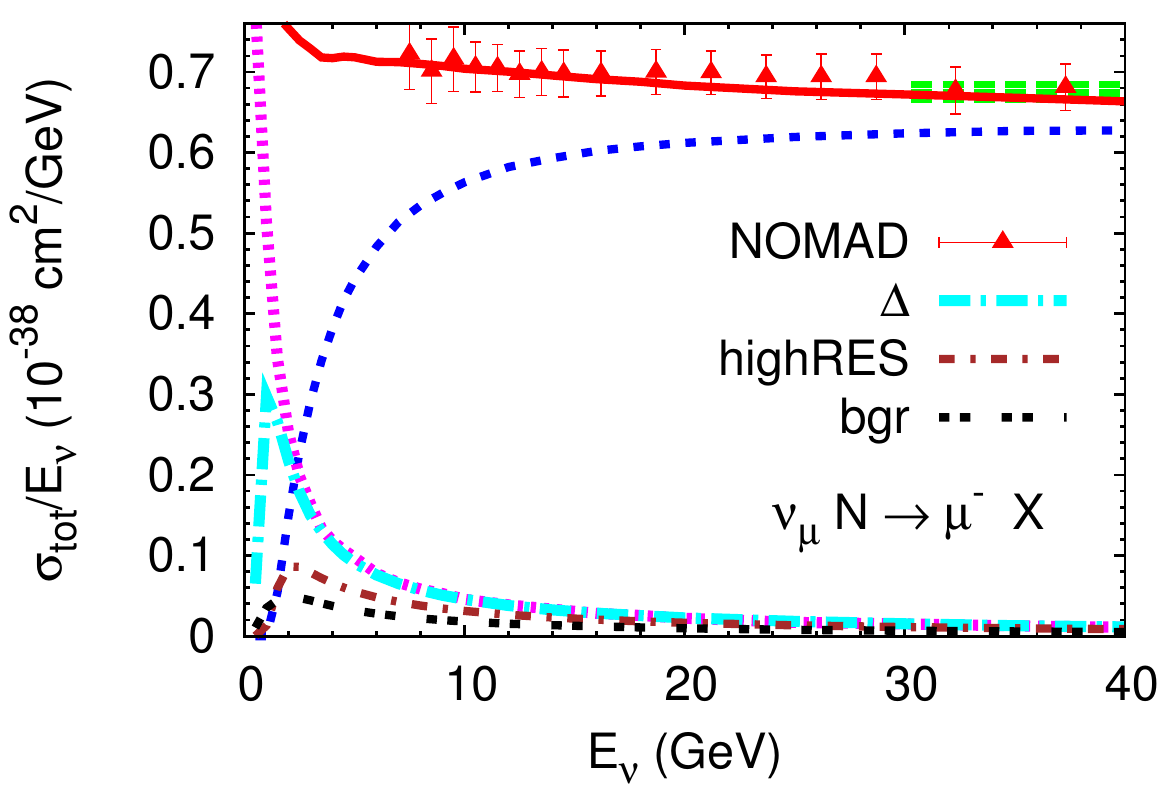}
\end{tabular}
\caption{Total (red solid), DIS (blue dashed) and other cross sections per energy per nucleon
for antineutrino (top panel) and neutrino (bottom panel)
inclusive scattering on an isoscalar target. Curves are compared with the world averaged values, denoted by the dashed band.
Data points from the NOMAD experiment \cite{:2007rv} are shown as solid triangles.}
\label{fig:compare-isoscalar-free}
\end{figure}

As we learned from photoproduction described in the previous section, we expect all
reaction channels to be important.
Thus, the total absorption cross section is then calculated as
$\sigma_{\rm tot}=\sigma_{\rm QE}+\sigma_{\rm RES}+\sigma_{\rm bgr}+\sigma_{\rm DIS}$.
Each term is shown explicitly in Fig.~\ref{fig:isoscalar-dsidQ2-dsidnu} for the example of
the $\dd\sigma/\dd Q^2$ and the $\dd\sigma/\dd \nu$ cross sections.
At low $Q^2$ all contributions are important. All of them except DIS, however, strongly fall off
with increasing $Q^2$. The DIS contribution, on the other hand, is much flatter in $Q^2$;
its absolute value will also grow  with increasing energy.
The dependence on energy transfer $\nu$ is dominated by different channels in different regions of $\nu$.
It is clearly seen, that as soon as the neutrino energy is large enough to allow $\nu>2\GeV$,
which is the case for the present experiments, all the processes will be important.

From Fig.~\ref{fig:isoscalar-dsidQ2-dsidnu} it is also clear, that
at neutrino energies above a few GeV the major contribution to the total cross section comes from DIS.
According to the predictions of the parton model, the DIS cross section grows linearly with energy.
At high neutrino energies the data are, therefore, conveniently presented as cross section
per energy $\sigma_\textrm{tot}/E_\nu$.

Figure~\ref{fig:compare-isoscalar-free} shows the results of our calculations of the total cross section
as well as the DIS and other contributions indicated in the figure for both neutrino and antineutrino scattering
on an isoscalar target. The world average values and their error bands are indicated for comparison for $E_\nu > 30 \GeV$.
The NOMAD experiment has recently performed measurements on a composite target with a measured composition of $52.43\%$ protons
and $47.57\%$ neutrons \cite{:2007rv}. The measurements were then corrected for the non-isoscalarity, and the results are
presented as isoscalar cross section. The data points are shown in Fig.~\ref{fig:compare-isoscalar-free} as solid triangles.

For neutrinos, the DIS contribution becomes larger than the $\Delta$ contribution
at about 3 GeV, and at 5 GeV it is already about $60\%$ of the total and reaches $95\%$ at higher energies.
The rest is to be attributed to other channels.
For antineutrinos,
the DIS contribution becomes larger than the $\Delta$ channel at about 4 GeV; it is $40\%$ of the total at $E_{\bar\nu}\sim 5\GeV$ and reaches $95\%$ at higher neutrino energies. This clearly has
implications for theoretical analyses of the cross section measurements in the MINOS and NO$\nu$A experiments.

Such interplay of various reaction mechanism is the distinguishing feature of the Shallow Inelastic
Scattering (SIS), which reveals itself for a wide range of neutrino energies of few GeV.

The dip in the total antineutrino cross section at $E_\nu\sim 3-4 \GeV$
(there is an indication for a similar effect in the neutrino cross section)
shows up as such only in the $1/E_\nu$ scaled cross section.
It is caused by an interplay of the downfall of the resonance contributions and the rise of DIS and could indicate that our model misses some strength here.
This intermediate energy region would be most sensitive to the pion (one or more) background.
A clarification must wait until much more precise data for $1\pi$ and $2\pi$ production become available.

\subsection{Nuclear effects in DIS}

\begin{figure*}
\begin{minipage}[c]{0.32\textwidth}
\includegraphics[width=\textwidth]{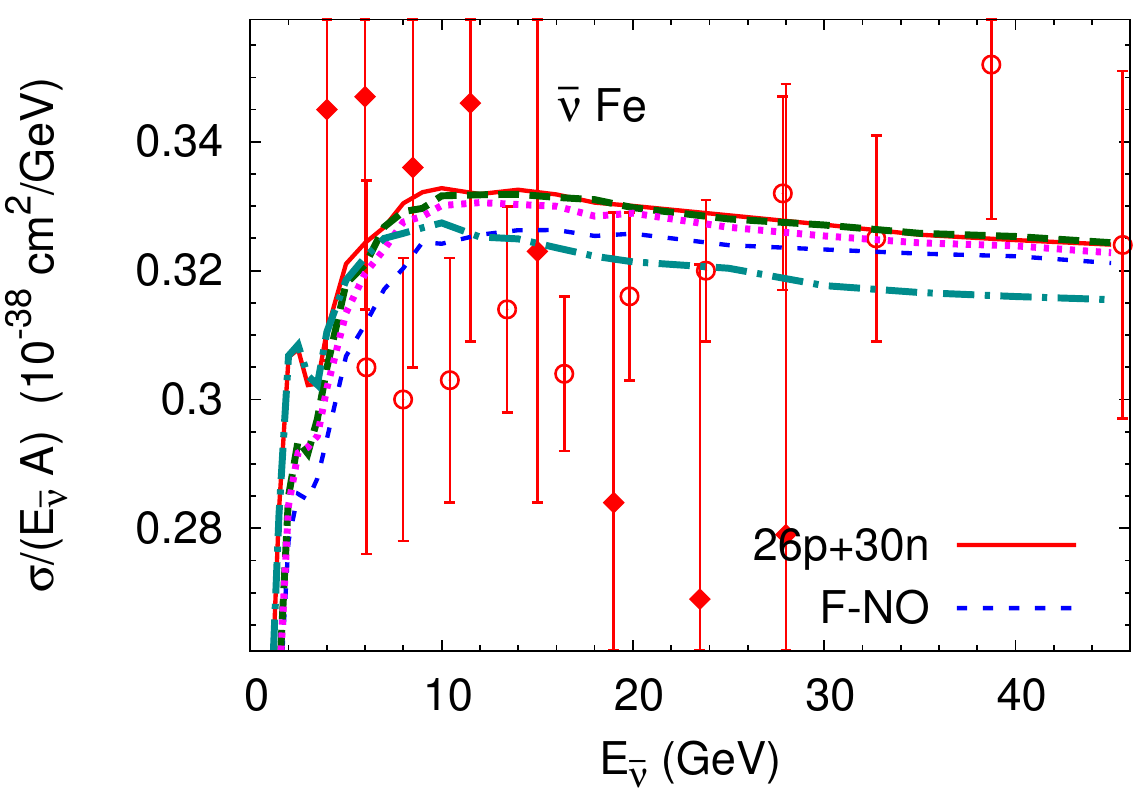}
\end{minipage}
\hfill
\begin{minipage}[c]{0.32\textwidth}
\includegraphics[width=\textwidth]{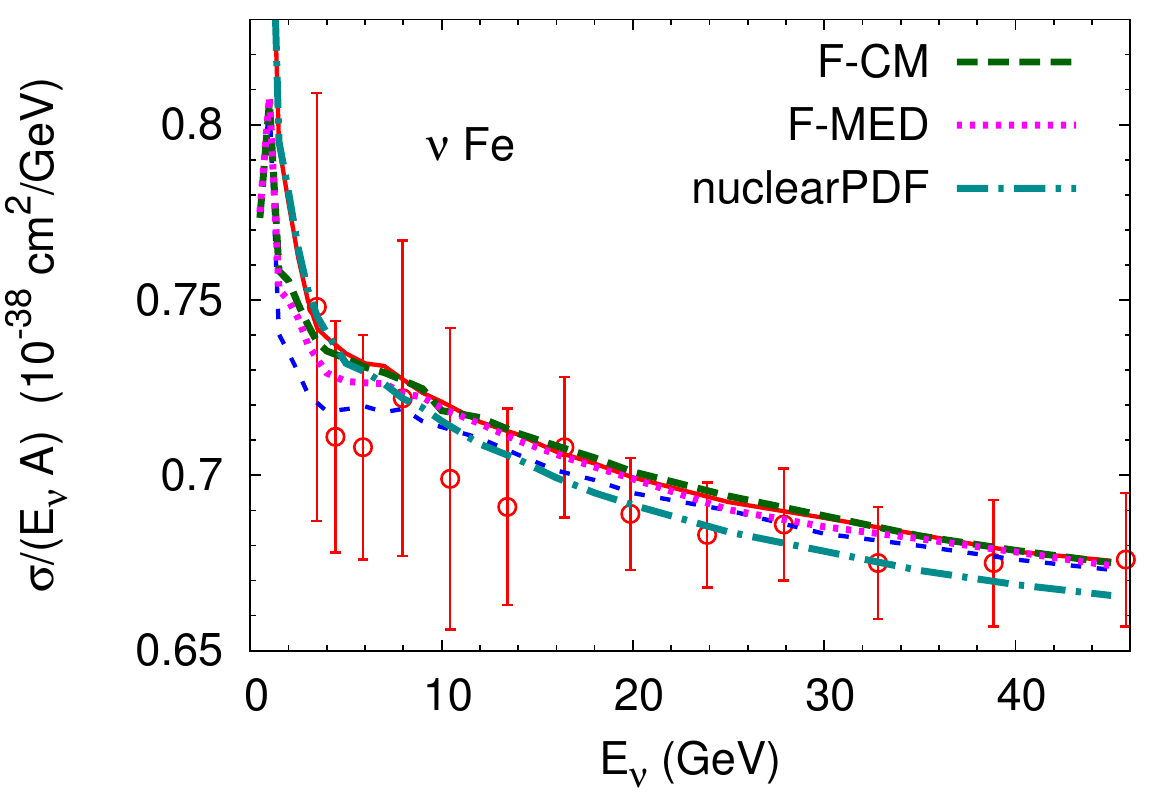}
\end{minipage}
\hfill
\begin{minipage}[c]{0.32\textwidth}
\includegraphics[width=\textwidth]{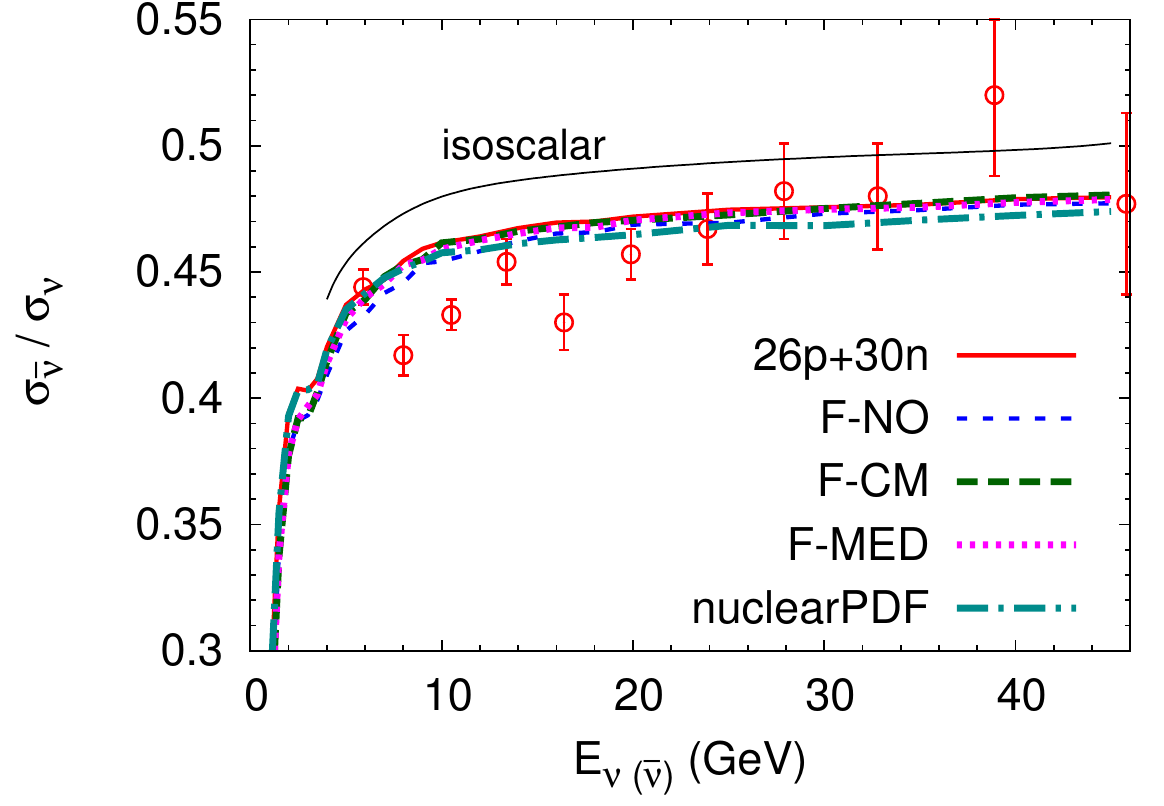}
\end{minipage}
\caption{Cross sections $\sigma_{\rm tot}/E_\nu$ per nucleon for antineutrino
$\bar\nu_\mu \mbox{Fe} \to \mu^+ X$ (left panel)
and neutrino $\nu_\mu \mbox{Fe} \to \mu^- X$ (middle panel) inclusive scattering off iron,
as well as  their ratio (right panel).
Also shown is the calculated ratio for an isoscalar-corrected target.
Experimental data are taken from Refs.~\protect\cite{Adamson:2009ju} (MINOS, open circles) and
\protect\cite{Anikeev:1995dj} (IHEP-JINR, solid diamonds).}
\label{fig:compare-nuclear}
\end{figure*}

Modern neutrino experiments often use the world-average inclusive neutrino cross sections
above neutrino energy of $30\GeV$ (also shown here in Fig.~\ref{fig:compare-isoscalar-free})
as a benchmark for normalization.
Recall, however, that all the old experiments, used to derive this average, were actually performed
on nuclear targets. If the target was non-isoscalar, a corresponding correction
was applied and the  isoscalar cross section was extracted \cite{Conrad:1997ne}.
Such a procedure as well as the introduction of the ``world-average'' value is meaningful
only if nuclear corrections are very small.

The EMC effect shows that there are nuclear corrections to the free cross sections for
electrons \cite{Piller:1999wx}.
For neutrino reactions, however, the situation is controversial.
On one hand, nuclear parton distributions, based on electromagnetic scattering data and intended for description
of both charged lepton and neutrino reactions, were introduced. For a review and a list
of recent parametrization see, for example, Ref.~\cite{Hirai:2009mq}.
On the other hand, a recent investigation \cite{Schienbein:2007fs,Kovarik:2010uv} showed that in neutrino reactions
nuclear corrections to parton distributions have about the same magnitude as for electrons, but
have a very different dependence on the Bjorken-$x$ variable.
The topic remains controversial \cite{Paukkunen:2013grz}, with the hope that future precise Miner$\nu$a results on various
targets will clarify the situation.

As we have already mentioned above, the GiBUU code uses \textsc{pythia} for the simulation of the DIS processes.
Since the \textsc{pythia} code was designed for elementary reactions
we have to provide some ``quasi-free'' kinematics as input to \textsc{pythia} that removes the effects of the binding potential on the nucleon.
Various prescriptions to do this  have been used (for details see Ref.~\cite{Buss:2011mx}) and are
compared with each other and with the free cross section in Fig.~\ref{fig:compare-nuclear}.
The corresponding cross sections are denoted as ``F-NO'' (the invariant energy $W^2$ of the boson-nucleon system
is calculated as $(k-k'+p)^2$ and not corrected),
``F-CM'' (bound nucleon is boosted into the center-of-momentum frame and the nuclear potential is removed
from its energy, the nucleon then is boosted back, and $W$ is calculated), and
``F-MED'' ($W^2$ is taken as  $(k-k'+p)^2 - m_N^*{}^2 + m_N^2$).
In all these calculations parton distributions appropriate for free nucleons have been used.

It is seen that the difference between these various prescriptions is quite small (about 2 \% at the lowest energy and less at the highest energy); also, the results all approach the free cross section at the highest energy. We consider small differences between the results obtained with the various prescriptions mentioned above as intrinsic uncertainty of the GiBUU code, reflecting the lack of a detailed understanding of nuclear effects. No other event generator, as far as we know, accounts for nuclear corrections in high--energy neutrino reactions. Nuclear parton distribution functions
from Ref.~\cite{Eskola:1998df} are also implemented as one of the options to use.
To avoid double counting, nuclear 35potential and Fermi motion are switched off in such calculations.
The result (``nuclearPDF'') as well as the free cross section for iron composition
(``26p+30n'') are also shown in Fig.~\ref{fig:compare-nuclear}.

Fig.~\ref{fig:compare-nuclear} shows that for antineutrinos our curves are  within the spread
of the MINOS and IHEP-JINR data \cite{Anikeev:1995dj}.
The overall agreement of our calculations with the data is,  therefore, better than the
agreement of the data with each other.
At low energies, where the main contribution comes from the QE and resonance production,
nuclear effects are known to reduce the
neutrino and antineutrino cross section.
This is why,  at $E_{\bar\nu}<5\GeV$, the curves ``F-NO'', ``F-CM'', and ``F-MED'', that take into account
the nuclear effects explicitly,
lie noticeably lower than the ``26p+30n'' curve. At high energies, however, the curves converge
towards each other.
The ``nuclearPDF'' curve, which implies modification of the DIS channel only, coincides
with the ``26p+30n'' curve at low energies
and consistently deviates from it to lower values of the cross section at higher
energies where DIS dominates.
The peak and dip in the region $3-4\GeV$ in the free cross section have the same
origin as for the isoscalar cross section,
as discussed in the previous section. Nuclear effects, mainly Fermi motion, wash
out this structure so it is no longer visible in the nuclear
cross sections.

For neutrinos our curves are in good agreement with the recent MINOS experiment, they lie within
the errors of the data points.
For both neutrinos and antineutrinos nuclear effects are noticeable at low energies. For higher
energies, $E_{\nu}>5\GeV$, the curves ``F-NO''
``F-CM'' and ``F-MED'' all approach each other and the ``26p+30n'' curve, reflecting the expected
disappearance of nuclear effects with increasing energy.

\begin{figure*}[hbt]
\hfill
\includegraphics[width=0.45\textwidth]{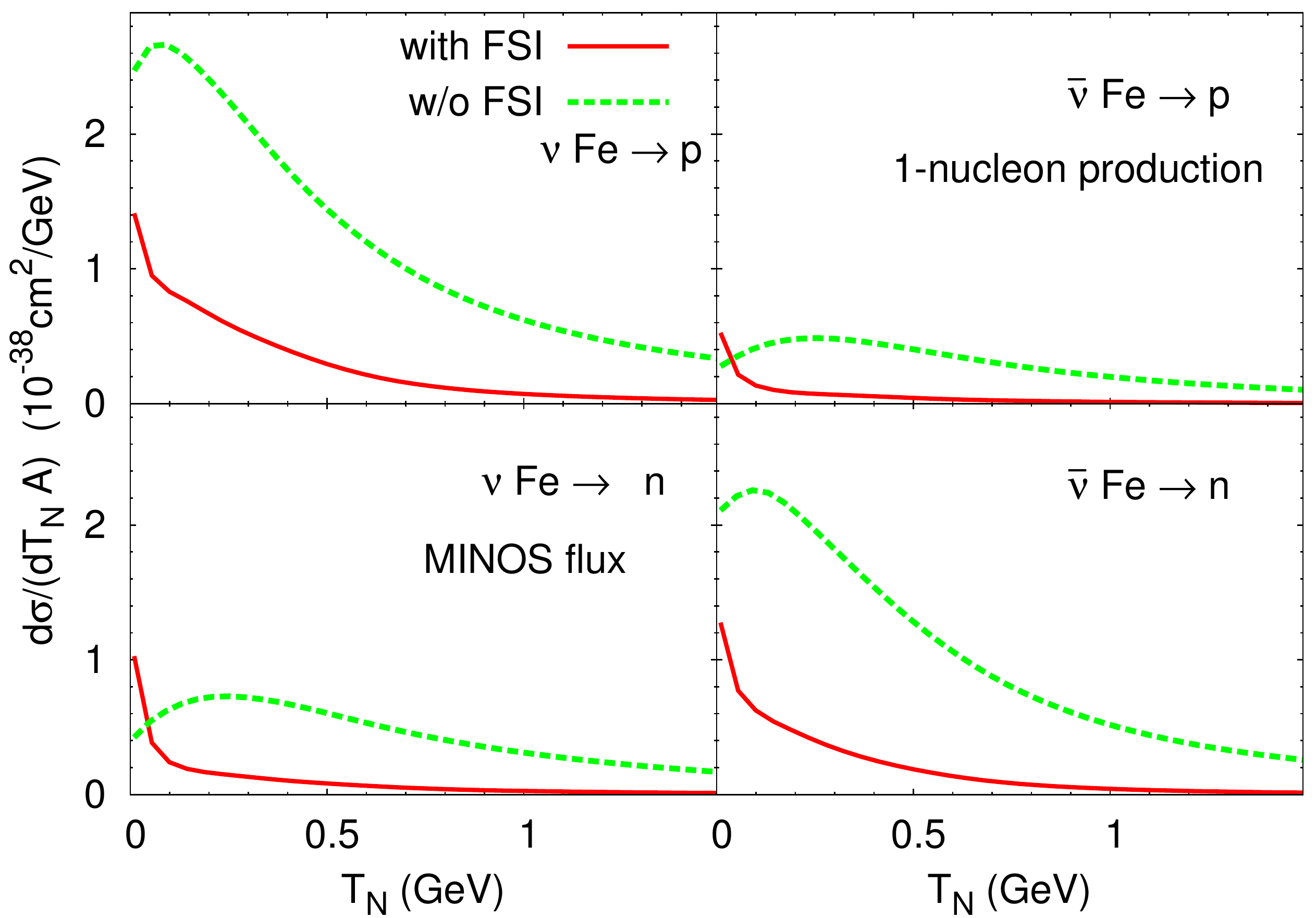}
\hfill
\includegraphics[width=0.25\textwidth]{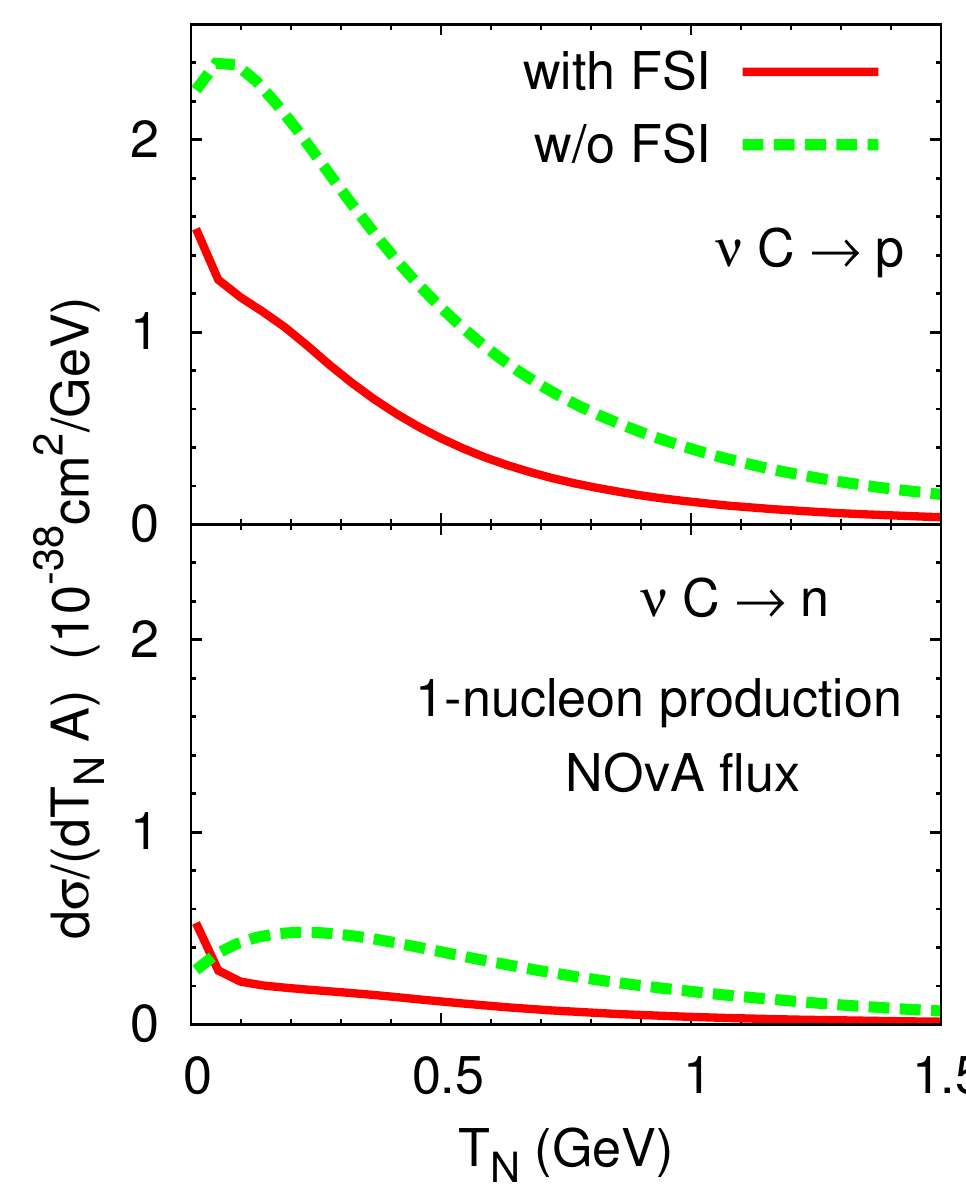}
\hfill
\caption{Kinetic energy distributions  per target nucleon  for \onenucleon
(one nucleon of the indicated charge and no other nucleons) production
in neutrino and antineutrino scattering off iron and carbon.
The left block of figures shows the results obtained for the MINOS flux on iron target,
and the right block those for the \Nova flux on carbon target. }
\label{fig:MINOS-ekin-with-wo-FSI-1-nucleon}
\end{figure*}
\begin{figure*}[bht]
\centering
\hfill
\includegraphics[width=0.45\textwidth]{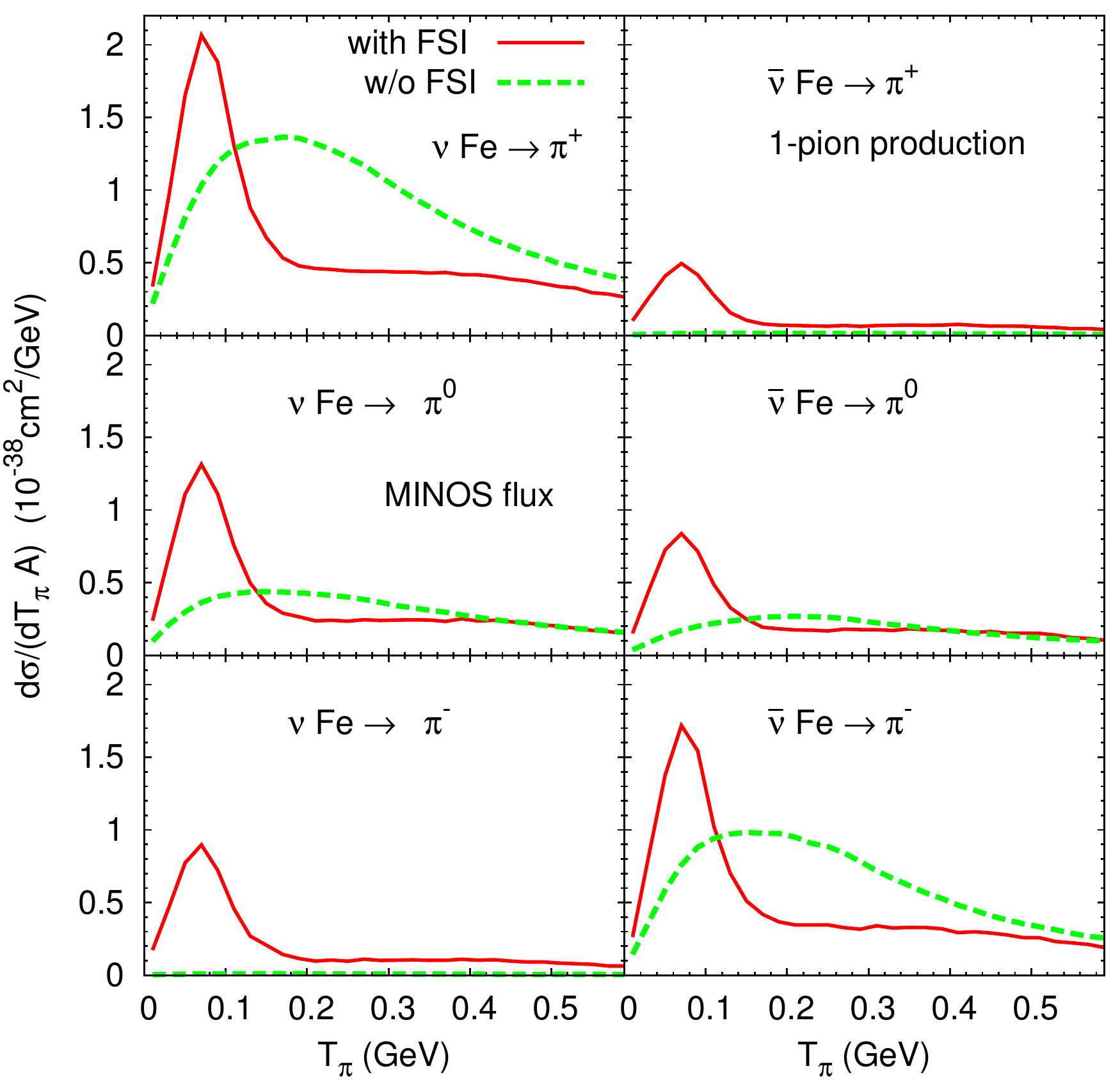}
\hfill
\includegraphics[width=0.25\textwidth]{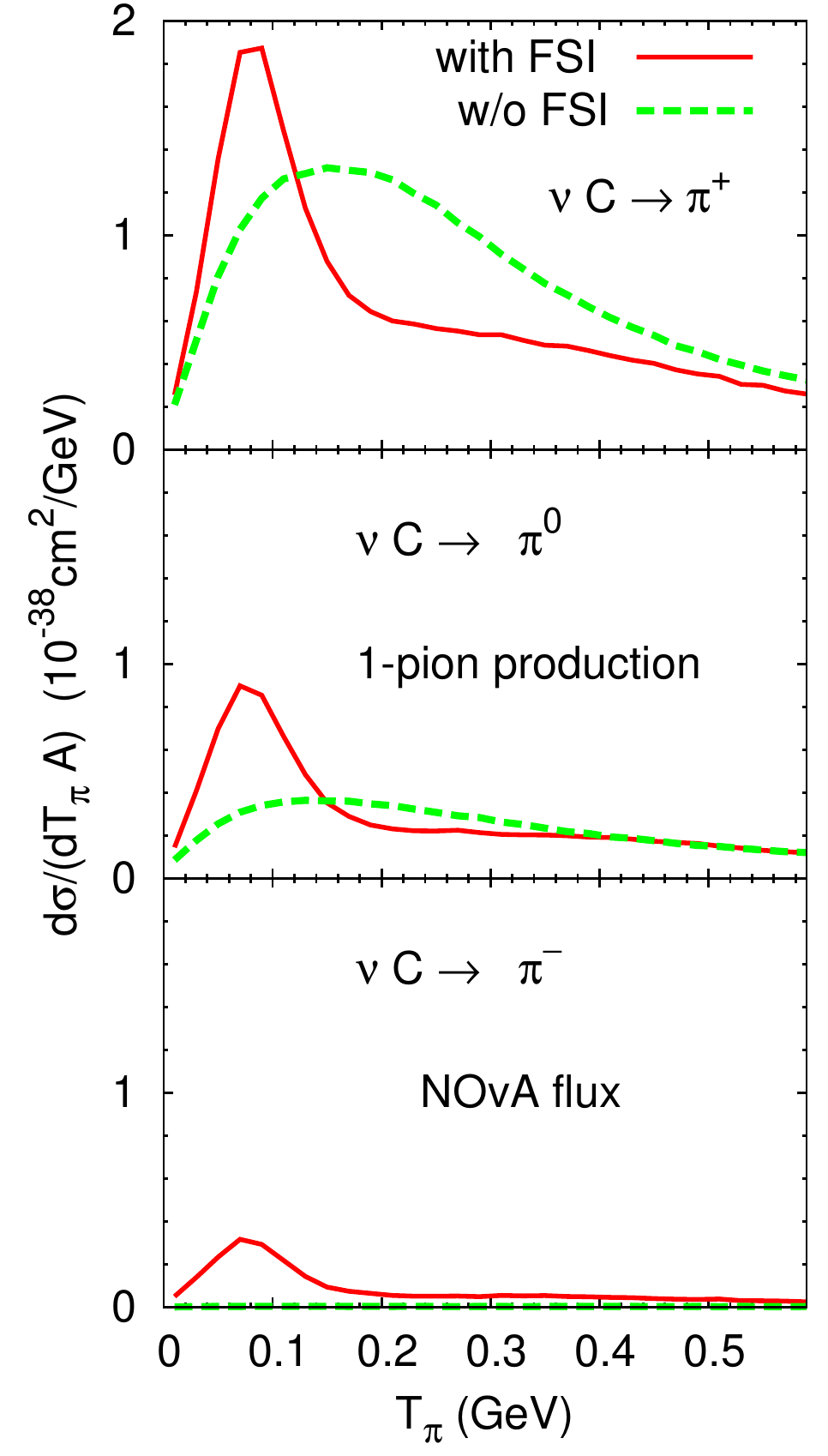}
\hfill
\caption{Pion kinetic energy distributions per target nucleon
for neutrino and antineutrino scattering off iron and carbon for \onepion production
(one pion of the indicated charge and no other pions are produced).}
\label{fig:MINOS-ekin-with-wo-FSI-1-pion}
\end{figure*}

In Fig.~\ref{fig:compare-nuclear} we also show the ratio of antineutrino cross section to the
neutrino one as a function of energy, which is in good agreement with the recent MINOS data.
The ratio rises with energy, gradually flattening out.  It is  below its asymptotic value obtained
for an isoscalar-corrected target ($\approx 0.5$ at high energies \cite{Conrad:1997ne});
the corresponding isoscalar  curve is also shown in the figure. It is interesting that this ratio
is remarkably insensitive to any nuclear effects, even at the lower energies, as illustrated by the
fact that now all the curves for the various in-medium correction methods lie essentially on top of
each other.

\subsection{Importance of final state interactions in semiinclusive reactions}

As we already mentioned, FSI can significantly change the experimental signature of
the initial neutrino-reaction and, thus, their realistic description is indispensable.
To demonstrate this point explicitly, in this section we present spectra for outgoing pions and nucleons for MINOS and \Nova experiments
before and after FSI.
No acceptance cuts and no detector thresholds are assumed for the outgoing hadrons.

The kinetic energy distributions of outgoing nucleons are shown in
Fig.~\ref{fig:MINOS-ekin-with-wo-FSI-1-nucleon}
for \onenucleon events (one nucleon of a given charge and no other nucleons in the final state).

One can easily see an essential decrease of the cross section after FSI as compared to those before FSI.
This decrease at higher kinetic energy energies ($T > 0.05\GeV$) is natural to expect,
because a nucleon can rescatter in the nucleus and knock out another nucleon;
the nucleon is then gone from the \onenucleon channel. At the same time, its kinetic energy would be
spread between the two secondary nucleons. If these two have an energy large enough, they could,
in turn, produce more lower-energy nucleons. The same knock-out can be caused by a pion produced
in a primary interaction. This process, which can develop as a cascade, leads to an increase of
multi-nucleon events with low kinetic energies.

Figure~\ref{fig:MINOS-ekin-with-wo-FSI-1-pion} shows the $\pi^+$, $\pi^0$
and $\pi^-$ spectra for  \onepion events
for neutrino- and antineutrino-induced reactions.
For the dominant channels ($\pi^+$ production in neutrino reactions and $\pi^-$ in antineutrino ones),
the FSI decrease the cross section at $T_\pi > 0.2 \GeV$. At lower neutrino energies this is mainly due to pion
absorption through $\pi N \to \Delta$ followed by $\Delta N \to NN$. At higher energies pions can also
be absorbed through $\eta$ and $\Delta$ production $\pi N\to\eta \Delta$,
production of higher resonances $\pi N \to R$ followed by $R N \to NN$ or $R\to \eta N$,
non-resonant pion absorption $\pi N N \to N N$, production of $\omega$ mesons
$\pi N \to \omega N$, $\phi$ mesons $\pi N \to \phi N$, and strange mesons
$\pi N \to \Sigma K, \, \Lambda K, \, K \bar{K} N$. All these channels (and more) are included in GiBUU and contribute
to pion absorption.

In addition, pion scattering in the FSI also decreases the pion energy.  Here elastic scattering as well as  DIS events of
the type $\pi N \to \mbox{multi-}\pi N$ deplete the spectra at higher energies and accumulate strength at lower energies.
Thus, an increase of the cross sections is observed at $T_\pi < 0.15 \GeV$, where the cross sections after FSI are higher than before,
and a decrease above this energy.
Additionally, low-energy pions may come from reactions such as $\eta N \to R$ followed by  $R \to \pi N$.
Altogether this leads to a significant change of the shape of the spectra.
In particular, there is a strong build-up of strength around $T_\pi = 0.06 \GeV$,
where the cross section after FSI is about 50\% higher than before.
This is primarily due to the slowing down of pions by FSI and the low $\pi-N$ cross section in
this energy region. We note that the size of this effect depends somewhat on the treatment of
the collisional width of the $\Delta$ resonance \cite{Leitner:2009zz}.

\begin{figure*}[h]
\centering
\hfill
\includegraphics[width=0.45\textwidth]{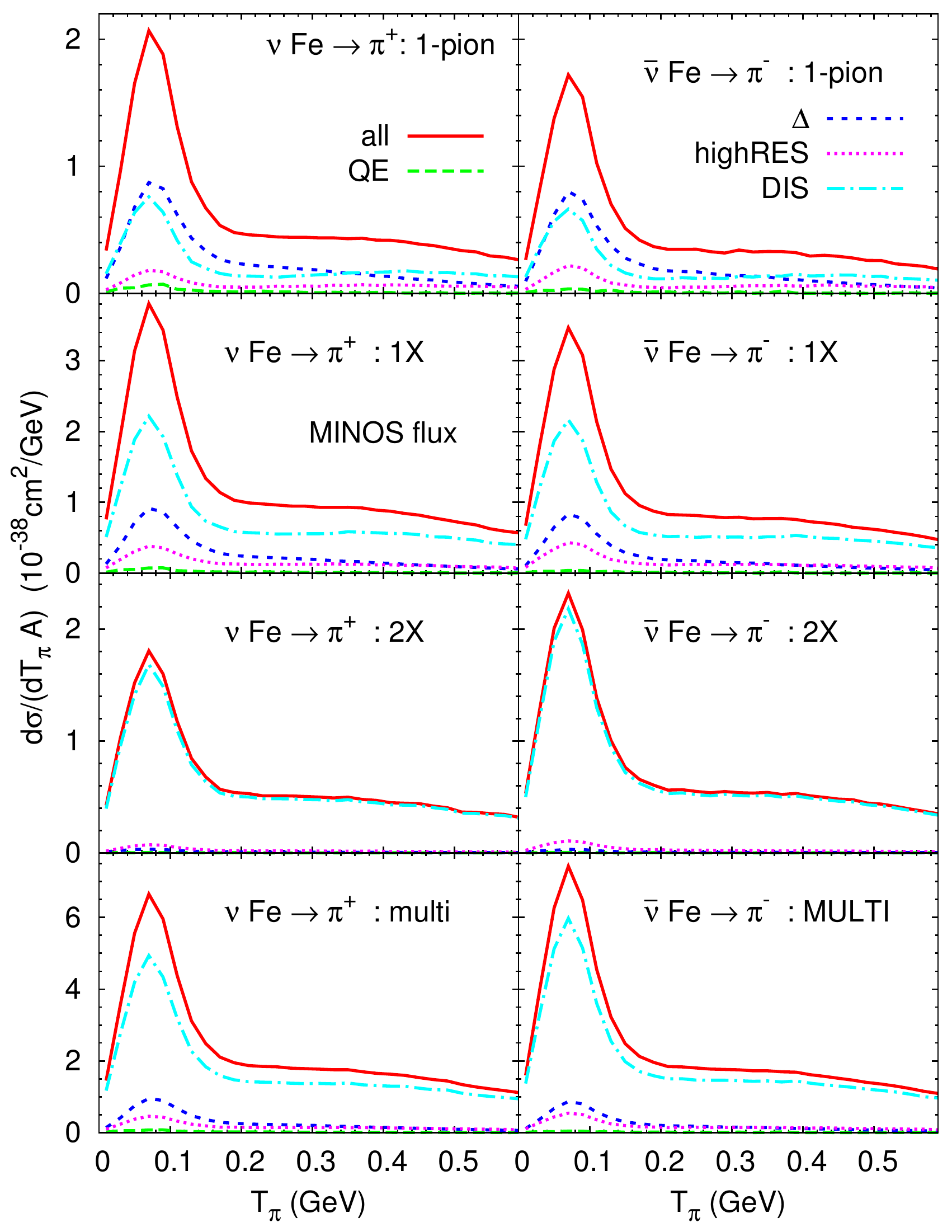}
\hfill
\includegraphics[width=0.26\textwidth]{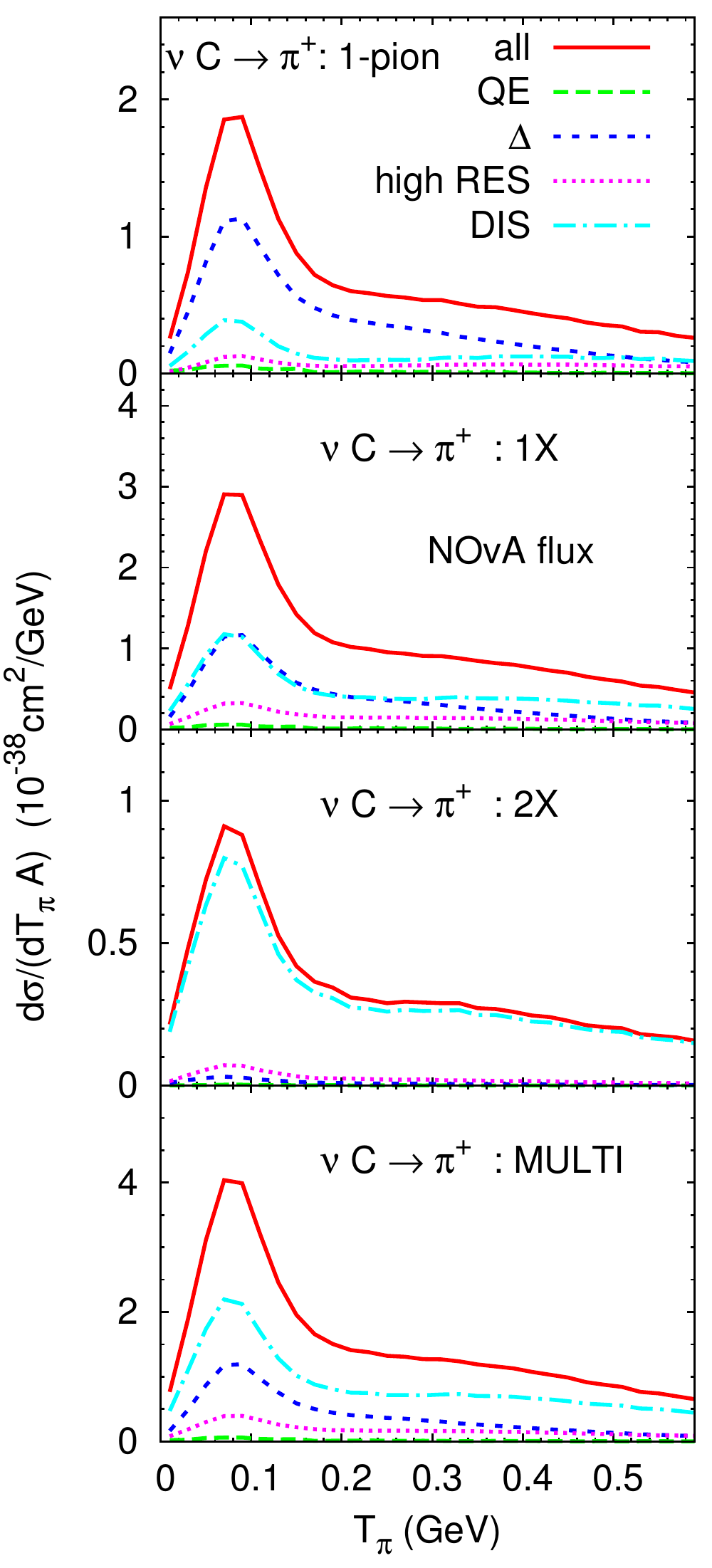}
\hfill
\caption{Pion kinetic energy distributions  per target nucleon for $\pi^+$
production in neutrino  and $\pi^-$ production in antineutrino scattering off iron and carbon
(both are dominant channels), showing various  contribution to a given final state:
''1-pion`` only one pion of the indicated charge and no other pions are produced; ''1X`` one pion of the indicated charge and
any number of pions of other charges are produced;  ''2X`` two pion of the indicated charge and
any number of pions of other charges are produced;  ``MULTI'' at least one pion of the indicated charge and any number
of pions of other charges are produced.}
\label{fig:MINOS-ekin-variousOrigins-nu-piplus-barnu-piminus}
\end{figure*}

The observed pions can be produced by different reaction mechanisms. It is,
therefore, interesting to analyze where the various final states come from.
Fig.~\ref{fig:MINOS-ekin-variousOrigins-nu-piplus-barnu-piminus} shows the origin of the pions
(that is, the initial vertex at which the pion was produced) in the dominant channels for various final states.
It is interesting to see that even for the MINOS flux, which peaks at $3\GeV$ and has a high-energy tail, \onepion production
receives its major contribution from $\Delta$ resonance production and
its following decay. The second largest contribution comes from DIS.
This reflects the fact that even in a DIS event the final state may involve a $\Delta$ that subsequently decays into a pion.
For the other final states with more than one pion DIS dominates, but the $\Delta$ is still visible.
This reflects the fact that the pion and nucleon produced in the decay of the primary $\Delta$ undergo  FSI
which result in several pions in the final state.
The contribution from the QE vertex is very small but nonzero. In this case the outgoing pion  can
be produced only during the FSI, for example, due to the $NN\to N\Delta$ scattering followed by $\Delta\to N\pi$.

Figs.\ \ref{fig:MINOS-ekin-with-wo-FSI-1-pion} and \ref{fig:MINOS-ekin-variousOrigins-nu-piplus-barnu-piminus} show that the pion spectra (after FSI) all exhibit
the typical shape that we already noticed for the calculated MiniBooNE pion spetra \cite{Lalakulich:2012cj}. This is so because
this spectral shape is a consequence of FSI alone (absorption of pions through the $\Delta$ resonance) and not of the
production mechanism. Typical for this higher energy is only the long tails towards higher pion energies which stems from DIS production
processes. The various production processes possible at these higher energies thus determine the absolute magnitude of the cross section, but not its shape.

We note here that the magnitude of the experimentally measured cross section depends on the flux used to extract it from the event rate and this flux carries its own uncertainties. The uncertainty in the overall flux magnitude (but not in its shape!) could be removed by comparing ratios of pion production cross sections for different charge states, but this would not constrain the underlying theoretical description of electroweak pion production. More interesting could be a comparison of magnitudes of the $1\pi$ vs.\ $2\pi$ (or even $n\pi$) channels because the latter are predominantly produced by DIS processes (see Fig.\ \ref{fig:MINOS-ekin-variousOrigins-nu-piplus-barnu-piminus}).

\section{Conclusions}
In conclusion we summarize here a few essential points.

\begin{itemize}
 \item Any neutrino event generator should have all initial reaction mechanisms (quasi-elastic scattering,
 $\Delta$ production, production of higher resonances, background 1- and many-pion production, DIS) under
 control. This is especially important for experiments working in the SIS energy region
 (nearly all current experiments).
 \item To improve modeling in the SIS region, elementary inputs (that is, cross sections on nucleon targets)
 are needed, especially for pion production.
 \item Nuclear dynamics should be checked with electron- and photon-induced reactions. There are lots of relevant data out there.
 \item Multiplicities of ejected hadrons give a sensitive test for final state interactions.
 \item Pions and nucleons in the detector are in general not those produced in the initial
 interaction vertex, but those after final state interactions. This is why a realistic modeling of
 FSI should be an important issue in any generator.
\end{itemize}

This work has been supported by DFG and BMBF.

\bibliographystyle{aipproc}
\bibliography{nuclear}

\end{document}